\documentclass{article}
\usepackage{arxiv}
\usepackage[utf8]{inputenc} 
\usepackage[T1]{fontenc}    
\usepackage{hyperref}       
\usepackage{url}            
\usepackage{booktabs}       
\usepackage{amsfonts}       
\usepackage{nicefrac}       
\usepackage{microtype}      
\usepackage{lipsum}
\usepackage{graphicx}
\usepackage{booktabs}       
\usepackage{amsfonts}       
\usepackage{nicefrac}       
\usepackage{microtype}      
\usepackage{lipsum}
\usepackage{graphicx}
\usepackage{amsmath,amssymb}
\usepackage{dcolumn}
\usepackage{bm}
\usepackage{braket}
\usepackage[dvipsnames]{xcolor}
\usepackage{xcolor,colortbl}
\usepackage{makecell,booktabs}
\usepackage{graphicx}
\usepackage{dcolumn}
\usepackage{listings}
\usepackage{bm}
\usepackage{braket}
\usepackage[dvipsnames]{xcolor}
\usepackage{xcolor,colortbl}
\usepackage{braket}
\usepackage{amsmath,amssymb,amsfonts}
\usepackage{algorithmic}
\usepackage{textcomp}
\usepackage{url}
\usepackage{bm}
\usepackage{times}
\usepackage{epsfig}
\usepackage{graphicx}
\usepackage{amsmath}
\usepackage{multirow}
\usepackage{amssymb}
\usepackage{subfig}
\usepackage{graphicx}
\usepackage{float}
\usepackage{booktabs}
\usepackage{textcomp}
\usepackage{adjustbox}\usepackage{mathtools}
\usepackage{array}
\usepackage{longtable}
\usepackage{makecell,booktabs}
\usepackage{multirow}
\usepackage{stackengine}
\usepackage{algorithm}
\usepackage{graphicx}
\usepackage{listings}
\usepackage{color}
\usepackage{color,array}
\usepackage{cite}
\usepackage{amsmath,amssymb,amsfonts}
\usepackage{algorithmic}
\usepackage{textcomp}
\usepackage{url}
\usepackage{bm}
\usepackage{times}
\usepackage{epsfig}
\usepackage{graphicx}
\usepackage{amsmath}
\usepackage{multirow}
\usepackage{amssymb}
\usepackage{subfig}
\usepackage{graphicx}
\usepackage{float}
\usepackage{booktabs}
\usepackage{textcomp}
\usepackage{adjustbox}\usepackage{mathtools}
\usepackage{array}
\usepackage{makecell,booktabs}
\usepackage{multirow}
\usepackage{stackengine}
\usepackage{algorithm}
\usepackage{graphicx}
\usepackage[dvipsnames]{xcolor}
\usepackage{xcolor,colortbl}
\newtheorem{definition}{Definition}
\graphicspath{ {./images/} }
\usepackage[title]{appendix}
\usepackage{hyperref}
\hypersetup{
    colorlinks=true,
    linkcolor=blue,
    filecolor=magenta,      
    urlcolor=cyan,
    pdftitle={Overleaf Example},
    pdfpagemode=FullScreen,
    }
\usepackage{setspace} 
\usepackage{lineno}
\definecolor{amber}{rgb}{1.0, 0.75, 0.0}
\definecolor{orange}{rgb}{1.0, 0.49, 0.0}
\definecolor{codegreen}{rgb}{0,0.6,0}
\definecolor{codegray}{rgb}{0.5,0.5,0.5}
\definecolor{codepurple}{rgb}{0.58,0,0.82}
\definecolor{backcolour}{rgb}{0.95,0.95,0.92}

\lstdefinestyle{mystyle}{
    backgroundcolor=\color{backcolour},   
    commentstyle=\color{codegreen},
    keywordstyle=\color{magenta},
    numberstyle=\tiny\color{codegray},
    stringstyle=\color{codepurple},
    basicstyle=\ttfamily\footnotesize,
    breakatwhitespace=false,         
    breaklines=true,                 
    captionpos=b,                    
    keepspaces=true,                 
    numbers=left,                    
    numbersep=5pt,                  
    showspaces=false,                
    showstringspaces=false,
    showtabs=false,                  
    tabsize=2
}
\renewcommand{\figurename}{Fig.}

\title{Biomarker Discovery with Quantum Neural Networks: A Case-study in \textit{CTLA4}-Activation Pathways}

\author{
    Phuong-Nam Nguyen \\
    Faculty of Computer Science\\
    PHENIKAA University, Yen Nghia, Ha Dong\\
    Hanoi, Vietnam 12116 \\
    \texttt{nam.nguyenphuong@phenikaa-uni.edu.vn} \\
}

\begin{document}
\maketitle
\begin{abstract}
    \textbf{Background:} Biomarker discovery is a challenging task due to the massive search space. Quantum computing and quantum Artificial Intelligence (quantum AI) can be used to address the computational problem of biomarker discovery from genetic data.\\
    \textbf{Method:} We propose a Quantum Neural Networks (QNNs) architecture to discover genetic biomarkers for input activation pathways. The Maximum Relevance — Minimum Redundancy (mRMR) criteria score biomarker candidate sets. Our proposed model is economical since the neural solution can be delivered on constrained hardware.\\
    \textbf{Results:} We demonstrate the proof of concept on four activation pathways associated with \textit{CTLA4}, including (1) \textit{CTLA4}-activation stand-alone, (2) \textit{CTLA4-CD8A-CD8B} co-activation, (3) \textit{CTLA4-CD2} co-activation, and (4) \textit{CTLA4-CD2-CD48-CD53-CD58-CD84} co-activation.\\ 
    \textbf{Conclusion:} The model indicates new genetic biomarkers associated with the mutational activation of \textit{CLTA4}-associated pathways, including 20 genes: \textit{CLIC4}, \textit{CPE}, \textit{ETS2}, \textit{FAM107A}, \textit{GPR116}, \textit{HYOU1}, \textit{LCN2}, \textit{MACF1}, \textit{MT1G}, \textit{NAPA}, \textit{NDUFS5}, \textit{PAK1}, \textit{PFN1}, \textit{PGAP3}, \textit{PPM1G}, \textit{PSMD8}, \textit{RNF213}, \textit{SLC25A3}, \textit{UBA1}, and \textit{WLS}. We open source the implementation at: \textcolor{blue}{\url{https://github.com/namnguyen0510/Biomarker-Discovery-with-Quantum-Neural-Networks}}.\\
\end{abstract}

\section{Introduction}\label{sec:intro}
A biomarker, a molecular marker or signature molecule, refers to a biological substance or characteristic found in body fluids, tissues, or blood that indicates the presence of a condition, disease, or abnormal process. Biomarkers can be measured to assess how well the body responds to treatment for a particular disease or condition\cite{nci-biomarker}. Biomarkers play a crucial role in drug discovery and development by providing essential information on the safety and effectiveness of drugs. These measurable indicators can be categorized into diagnostic, prognostic, or predictive biomarkers, and they are utilized to choose patients for clinical trials or track patient response and treatment efficacy. The Next-Generation Sequencing (NGS) technology has revolutionized the field of oncology by enabling the comprehensive and precise identification of \textit{genetic biomarkers}, paving the way for personalized cancer therapies and improved patient outcomes.

\textcolor{blue}{The study\cite{degroat2023intelligenes} introduces a new metric, the Intelligent Gene (I-Gene) score, to measure the importance of individual biomarkers for predicting complex traits. Their Machine learning (ML) pipeline combines classical statistical methods and state-of-the-art algorithms for biomarker discovery. Another research group develops an autoencoder-based biomarker identification method by reversing the learning mechanism of the trained encoders\cite{al2022biomarker}. It provides an explainable post hoc methodology for identifying influential genes likely to become biomarkers. In \cite{fang2023deep}, a Deep learning (DL) pipeline predicts the status of five biomarkers in LGG using slide-level biomarker status labels and whole slide images stained with hematoxylin and eosin. The research assesses the performance of several state-of-the-art Random Forest (RF) based decision approaches, including the Boruta method, permutation-based feature selection with and without correction, and the backward elimination-based feature selection method\cite{acharjee2020random}. A review offers tips to overcome common challenges in biomarker signature development, including supervised and unsupervised ML, feature selection, and hypothesis testing\cite{diaz2022ten}. Another systematic review examines the current state of the art and computational methods, including feature selection strategies, ML and DL approaches, and accessible tools to uncover markers in single and multi-omics data\cite{dhillon2023systematic}.} Despite its valuable role, genetic biomarker discovery is a challenging task for classical-computational platforms due to the massive search space (\textbf{Section}~\ref{sec:problem_state}). 

Quantum computing is an emerging technology that utilizes the principles of quantum mechanics to solve problems beyond classical computers' capabilities. Quantum Machine Learning and Quantum Neural Networks are an advanced class of machine intelligence on quantum hardware, which promises more powerful models for myriad learning tasks (\textbf{Section}~\ref{sec:qnn}). \textcolor{blue}{A comprehensive review discusses quantum computing technology and its status in solving molecular biology problems, especially in the next-generation computational biology scenario\cite{pal2024quantum}. The review covers the basic concept of quantum computing, the functioning of quantum systems, quantum computing components, and quantum algorithms. HypaCADD, a hybrid classical-quantum workflow for finding ligands binding to proteins, is introduced in \cite{lau2023insights}. While accounting for genetic mutations, it combines classical docking and molecular dynamics with QML to infer the impact of mutations. The study found that the QML models can perform on par with, if not better than, classical baselines. Another systematic review presents the recent progress in quantum computing and simulation within the field of biological sciences. It discusses quantum computing components, such as quantum hardware, quantum processors, quantum annealing, and quantum algorithms\cite{pal2023future}. A review comments on recently developed Quantum Computing (QC) bio-computing algorithms, focusing on multi-scale modeling and genomic analyses\cite{marchetti2022quantum}. The research group highlights the possible advantages over the classical counterparts and describes some hybrid classical/quantum approaches. In a non-conventional track, scientists at Oak Ridge National Laboratory used their expertise in quantum biology, artificial intelligence, and bioengineering to improve how CRISPR Cas9 genome editing tools work on organisms like microbes that can be modified to produce renewable fuels and chemicals\cite{ornl2023}. The cross-section of the genetic editing technology with quantum methods shows promising new insights into biomedical science.}

This work uses a class of Quantum Artificial Intelligence (AI) models to discover genetic biomarkers in biomedical research. We adopt the neural architecture proposed in a recent work that addresses the body dynamics modeling problem\cite{nguyen2024duality}. Here, we make a non-trivial adaptation of the proposed game theory to tackle a different class of problems. The main contribution of this study is summarized as follows:
\begin{enumerate}
    \item The proposed quantum AI model is a general, cost-efficient, cost-effective algorithm for biomarker discovery from genetic data despite the extensive problem complexity.
    \item The model outcomes suggest novel biomarkers for the mutational activation of the notable target in immuno-therapy - \textit{CLTA4}, including $20$ genes: \textit{CLIC4}, \textit{CPE}, \textit{ETS2}, \textit{FAM107A}, \textit{GPR116}, \textit{HYOU1}, \textit{LCN2}, \textit{MACF1}, \textit{MT1G}, \textit{NAPA}, \textit{NDUFS5}, \textit{PAK1}, \textit{PFN1}, \textit{PGAP3}, \textit{PPM1G}, \textit{PSMD8}, \textit{RNF213}, \textit{SLC25A3}, \textit{UBA1} and \textit{WLS}.
\end{enumerate}
We organize the article as follows: Section~\ref{sec:prelimns} formalizes the biomarker identification problem as a combinatory optimization problem and the preliminary for QNN models; Section~\ref{sec:method} introduces our proposed model architecture and the scoring algorithm; Section~\ref{sec:results} reports the in \textit{silico} discovery for genetic biomarkers of four immunotherapy pathways with posthoc validation using literature mining over clinical research; Section~\ref{sec:conclusion} concludes our research by suggesting several further research direction.

\section{Preliminary}\label{sec:prelimns}
\subsection{Problem Statement}\label{sec:problem_state}
The learning task involves identifying the best combinations of genetic biomarkers from a given set of genes (represented as $\mathcal{G}$) to select those that are (1) most relevant to a specific pathway (represented as $\bm{Y}$) and (2) optimal for machine learning algorithms. This criterion is commonly referred to as minimizing redundancy and maximizing relevancy for selected feature sets, as proposed in\cite{peng2005feature} (See \textbf{Appendix}~\ref{app:mrmr}).

In this context, it is worth noting that each individual has unique patterns of mutational alterations that lead to distinct sets of genetic biomarkers. Considering all the possibilities, the number of candidate biomarker sets is the sum of all possible combinations, which can be expressed as 
\begin{equation}
    \sum_{i=0}^{N}{N\choose k} = 2^N   
\end{equation}
Since the human genome contains around 20,000 to 25,000 genes, this results in a massive search space of $3.2019 \times 10^{6577}$ candidate biomarker set for any biomarker identification algorithm from genetic databases. This search space grows exponentially with the number of input genes, making it an incredibly challenging problem for conventional computing methods.
\textcolor{purple}{Quantum computing holds great promise for advancing genomic research, particularly in quantifying biomarkers. In this context, the logarithmic scaling complexity of quantum algorithms becomes evident, as exemplified by the requirement of \(\log_2{20,000}\) to \(\log_2{25,000}\) noise-tolerant qubits for quantifying genome sets, approximately equivalent to 15 qubits. The salient advantage of employing quantum hardware for biomarker quantification lies in the \(O(\log N)\) scaling complexity, providing a significant computational advantage as the genomic dataset size increases. For instance, in the presence of four multimodality datasets encompassing DNA Methylation, RNA, mRNA, and Protein, each with a homogeneous number of features denoted as \(M\), the required number of qubits scales only to \(N = \log_2{4M} = 2 + \log_2{M}\), illustrating a scalable problem complexity. However, the current state-of-the-art quantum hardware faces limitations in facilitating multimodality analysis, primarily due to the constraints imposed by the limited and noisy qubits available, hindering their effectiveness in realizing the full potential of quantum computational power in genomics.}

\subsection{Quantum Neural Networks}\label{sec:qnn}
QNNs represent input data using wavefunction representations, typically using qubits or "qurons"\cite{schuld2019quantum}, in the form of:
\begin{equation}
    \ket{\psi} = \alpha\ket{0} + \beta\ket{1}, \alpha \text{ and } \beta \in \mathbb{C}.
\end{equation}
This distinguishes QNNs from classical ANNs as they capture physical events discretely\cite{mcculloch1943logical} and offer insight into representation learning\cite{bengio2013representation}.

QNNs have been proposed to offer two advantages over classical ANNs\cite{schuld2022quantum}. Quantum feature maps can represent exponentially larger data sets than classical neural networks, with an $n$-qubit system representing $2^n$ bits\cite{nielsen2002quantum}. Secondly, quantum feature maps inherit wavefunction uncertainty from quantum mechanics, allowing measurements of quantum states to return values of $0$ or $1$ with probability values of $\mathbb{P}(0) = |\alpha|^2$ and $\mathbb{P}(1) = |\beta|^2$. In our previous papers, we discuss the role of epistemic uncertainty estimation from quantum maps\cite{nguyen2022bayesian} and the effect of entanglement layouts on the classifier's performance.

The typical design for QNNs\cite{farhi2018classification,romero2021variational,benedetti2019parameterized,schuld2021effect,killoran2019continuous} involves a stack of identical ansatz structures, given by a global unitary transformation:
\begin{equation}
    U_{\bm{\theta}} (\bm{x}) = \mathcal{U}^{(l)}(\bm{\theta}l)\mathcal{V}(\bm{x}) \dots \mathcal{V}(\bm{x}) \mathcal{U}^{(0)}(\bm{\theta}{0}),
\end{equation}
where $\bm{x}$ represents input features, $\bm{\theta}$ represents model weights, $\mathcal{U}^{(l)}(.)$ represents identical-parameterized circuits serving as variational (trainable neural) blocks, and $\mathcal{V}(.)$ represents feature-embedding blocks. From our perspective, QNNs are advanced mathematical models in the language of representation theory (See \textbf{Appendix}~\ref{app:qnn}); thus, we are using mathematics to enable cancer discoveries.

\section{Method}\label{sec:method}
\begin{figure}
    \centering
    \includegraphics[width  = \textwidth]{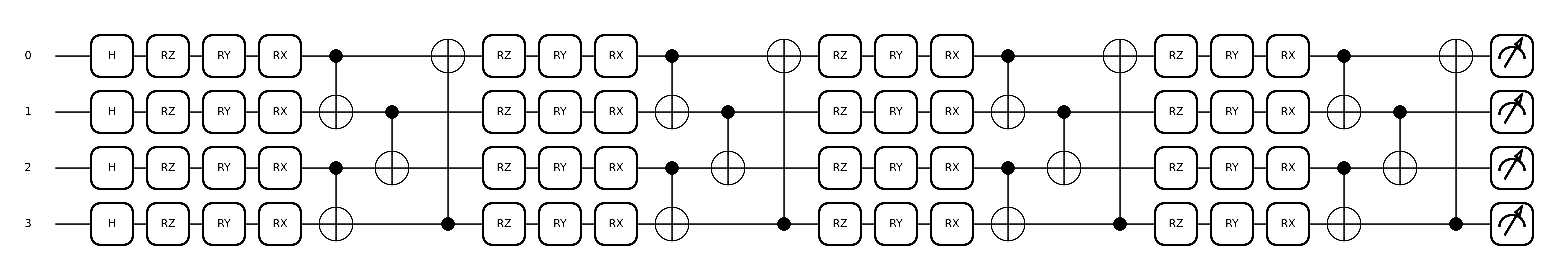}
    \caption{\textbf{The Ansatz Circuit of Our Proposed QNN Models using Four Qubits with Four Neural Blocks}. Here, the number of evaluated genes is $2^4 = 16$ genes. We extend the architecture to $11$-qubit ansatz in the numerical result, scoring $2^{11} = 2,048$ genes.}
    \label{fig:qnn_model}
\end{figure}
\subsection{Model Architecture}
We depict the ansatz structure of our QNN model in \textbf{Figure}~\ref{fig:qnn_model}. Each neural block in our proposed model includes (1) a parameterized RZ-rotation based on time-domain variables $t \in [0,2\pi]$, given by
\begin{equation}\label{equa:rz}
    \bm{R}_Z(t) = \begin{bmatrix}
    e^{-i\dfrac{t}{2}} & 0\\
    0 & e^{i\dfrac{t}{2}}
    \end{bmatrix}
\end{equation}
and (2) trainable RX-rotation and RY-rotation gates
\begin{equation}\label{equa:rx}
    \bm{R}_{X}(\alpha) = \begin{bmatrix}
        \cos(\alpha/2) & -i \sin(\alpha/2) \\
        -i \sin (\alpha/2) & \cos(\alpha/2)
    \end{bmatrix}
\end{equation}
and 
\begin{equation}\label{equa:ry}
    \bm{R}_{Y}(\beta) = \begin{bmatrix}
        \cos(\beta/2) & - \sin(\beta/2) \\
        \sin (\beta/2) & \cos(\beta/2)
    \end{bmatrix}
\end{equation}

The parameterized wavefunction of the 1-layer model is given
\begin{equation}
    \ket{\psi_{\Theta}} = \ket{\psi_{(\bm{\alpha}, \bm{\beta})}(\bm{t})} = \text{CNOT} (\Lambda)  \bm{R}_{X}(\bm{\alpha}) \bm{R}_{Y}(\bm{\beta}) \bm{R}_{Z}(\bm{t}) H \ket{0}^{\otimes n},
\end{equation}
$n$ is the number of qubits, and $\Lambda$ is the architecture parameter of the entanglement layout constructed by CNOT gates, illustrated in \textbf{Figure}~\ref{fig:qnn_model}. 

\subsection{Algorithm}
\subsubsection{Computations of Gene Scores (\textbf{GSCORE}$^\copyright$)}
We train the proposed QNNs to learn the optimal sampling distribution based on mRMR criteria (See \textbf{Appendix}~\ref{app:mrmr}). In other words, the search algorithm will assign higher probabilities for the more important markers. Thus, the search algorithm will more often select more important genetic biomarkers. Specifically, the output electronic wavefunction of the parameterized QML model is given by
\begin{equation}
    \ket{\psi_{\Theta}} = \mathcal{U}_{n}(\Theta),
\end{equation}
with $n$ is the number of prepared qubits and $\Theta$ is the model parameter. The sampled probability density is given by
\begin{equation}
    p(\Theta) = |\Psi(\Theta)|^2 = \Psi^{*} \Psi
\end{equation}
where $\ket{\Psi^{*}}$ is the conjugate-transpose of the output wavefunction. In other words, we compute the wavefunction's probability amplitude or square modulus.

We further use the Softmax function with tunable temperature to normalize our density. Besides, using the Softmax function also allows us to control the conservative of the search engine as the low temperature will encourage the model confidence. In contrast, high temperatures encourage less conservative predictions. Thus, the biomarker score is the sampling probability of each gene given by
\begin{equation}
    \bm{\text{GSCORE}}^\copyright = \hat{p}(\Theta) = \text{SoftMax} \bigg( \dfrac{\sqrt{p(\Theta)}}{\text{temp}} \bigg),
\end{equation}
where $\text{temp}$ is the function temperature. Finally, we select the candidate marker set by parameterized thresholding with $\bar{p} = 1$ if $\hat{p} \geq \tau$, otherwise $0$. We interpret the GSCORE$^\copyright$ that the more important genes have a higher chance to be selected by the sampler, resulting in a higher probability over the output of quantum ansatzes.

\subsubsection{Objective Function} 
We adopt the efficient loss function of the Quadratic Programming Feature Selection (QPFS) method\cite{rodriguez2010quadratic}, given as
\begin{equation}
    \min_{\bm{\Theta}} \bigg( \lambda \bm{p}^\intercal \bm{H} \bm{p} - \bm{p}\bm{F} \bigg), \sum_{i=1}^{n}p_i = 1, p_i > 0, 
\end{equation}
where $\bm{F}_{n\times 1}$ is the relevancy to target variables and $\bm{H}_{n\times n}$ is the pairwise redundancy computed from the feature set. Both natural and normalized quantum distributions $p(\Theta)$ and $\hat{p}(\Theta)$ satisfy the conditions for $p_i$ in QPFS. We consider $\lambda = 1$ for further analysis, i.e., the balanced loss between redundant-relevant criteria.

\subsubsection{Pseudo-code}\label{sec:pseudo-code}
We implement the proposed model using the quantum simulation package Pennylane\cite{bergholm2018pennylane} with Pytorch 3.7\cite{paszke2019pytorch}. The mutual information criteria are computed by Scikit-learn\cite{pedregosa2011scikit}, and model optimization is conducted by Optuna\cite{akiba2019optuna} with Tree Parzen Estimators\cite{bergstra2011algorithms}. The pseudo-code is given in the following algorithm:

\begin{lstlisting}[language=Python, caption= Model Architecture of The Duality Game Model Developed for Biomarker Identification Task.]{QNN Model}
import pennylane as qml
import torch
import numpy as np
q = 11 #Number of Qubits
dev = qml.device("default.qubit", wires=q)
device = torch.device("cuda:0" 
    if torch.cuda.is_available() else "CPU")
@qml.qnode(dev, interface="torch")
def quantum_sampler(a,b,n):
  t = torch.tensor(np.linspace(0,np.pi, n))
  for i in range(q):
      qml.Hadamard(wires=i)
  for k, dt in enumerate(t):
      for i in range(q):
          qml.RZ(dt,wires=i)
          qml.RY(a[k,i],wires=i)
          qml.RX(b[k,i],wires=i)
      for i in range(0, q - 1, 2):
          qml.CNOT(wires=[i, i + 1])
      for i in range(1, q - 1, 2):
          qml.CNOT(wires=[i, i + 1])
      qml.CNOT(wires = [q-1,0])
  return qml.state()
\end{lstlisting}
\begin{table}
    \centering
    \scalebox{0.9}{
    \begin{tabular}{|c|c|c|}
        \toprule
        \textbf{Parameters} & \textbf{Range} & \textbf{Role}\\
        \midrule
        Number of Qubits  & $11$& Sampler for $2^{11} = 2048$ genes\\
       $\bm{\alpha}$   & $[-2\pi, 2\pi]$& $\bm{R}_X$ rotation \\
       $\bm{\beta}$  & $[-2\pi, 2\pi]$ &  $\bm{R}_Y$ rotation\\
        $\epsilon$  & $[\pi/96,\pi/24]$& Standard Deviation of Model Weight\\
       $\tau$ & $[0.5, 0.7]$& Threshold Value\\
       $\text{temp}$ & $[1/10^3, 2\times 10^{3}]$ & SoftMax Temperature\\
       \bottomrule
    \end{tabular}}
    \caption{\textbf{Hyper-parameter of Our Model Optimization.}}
    \label{tab:hyper_params}
\end{table}

\subsubsection{Hyper-parameter and Training Protocol}
We report the hyper-parameters for our search engine powered by proposed QNNs in \textbf{Table}~\ref{tab:hyper_params}. Noteworthy, we adopt the Tree Parzen Estimator with Sequential Model-based Optimization\cite{bergstra2011algorithms} (SMBO) to train our model. We learned from our previous works - CSNAS\cite{nguyen2021csnas} and BayesianQNN\cite{nguyen2022bayesian} that the used optimization is a cost-efficient algorithm, which enables effective search on a massive search space.

\section{Results}\label{sec:results}
\subsection{Case-study}\label{sec:casestudy}
\subsubsection{\textit{CTLA4}-activation Pathways}\label{sec:casestudy_ctla4}
\textit{CTLA4} - The gene is part of the immunoglobulin superfamily and produces a protein that transmits a signal to inhibit T cells. Mutations in this gene have been linked to various autoimmune diseases, including insulin-dependent diabetes mellitus, Graves disease, Hashimoto thyroiditis, celiac disease, systemic lupus erythematosus, and thyroid-associated orbitopathy\cite{griffith2017civic}. \textcolor{purple}{\textit{CD8A} and \textit{CD8B} encode the alpha and beta chains of the CD8 antigen, respectively. The CD8 antigen is a cell surface glycoprotein found on most cytotoxic T lymphocytes that mediate efficient cell-cell interactions within the immune system\cite{borras2023single}. The CD8 antigen acts as a coreceptor with the T-cell receptor on the T lymphocyte to recognize antigens displayed by an antigen-presenting cell in class I MHC molecules. Moreover, CD8+ T cells play a significant role in the response to immunotherapies that target CTLA4\cite{kristensen2020monitoring}. Following a common preclinical combination treatment protocol, the study used a radiolabeled antibody to detect changes in CD8a+ infiltration in murine colon tumors. The results showed that the treatment effectively inhibited tumor growth and increased the overall survival of mice.
\textit{CD2} and \textit{CD28} are both important co-receptors involved in T-cell activation\cite{green2000coordinate,skaanland2019carboxyl}. They are part of the immunoglobulin superfamily and are known to regulate T-cell activation in a coordinated manner. \textit{CD2} enhances adhesion between T cells and target cells and delivers an activation signal\cite{feldhaus1997cd2}. On the other hand, \textit{CD28} is a costimulatory receptor that can strongly enhance TCR signaling responses. It is believed that \textit{CD28} and \textit{CD2} may function together to facilitate interactions of the T cell and antigen-presenting cells (APCs), allowing for efficient signal transduction through the TCR\cite{green2000coordinate}. The precise mechanisms of \textit{CTLA4}’s inhibitory role are not fully understood, but it is believed that \textit{CTLA4} can compete with \textit{CD28} for ligand binding, acting as an antagonist of \textit{CD28}-mediated costimulation\cite{rowshanravan2018ctla}. This interaction is thought to occur at the immune synapse between T cells and antigen-presenting cells (APCs), where \textit{CTLA4} has been shown to recruit \textit{CD80}, thereby limiting its interactions with \textit{CD28}. Studying the coactivation of \textit{CTLA4} and \textit{CD2} could provide valuable insights into T-cell activation and immune regulation. However, limited research addresses the coactivation of \textit{CTLA4} and \textit{CD2}. This suggests further investigation is needed to establish a connection and understand its implications for immunotherapy and autoimmune disease treatment. Understanding these interactions could potentially lead to the development of more effective therapeutic strategies.
In this study, we aim to broaden the scope of immunological research by studying the coactivation of \textit{CTLA4}, \textit{CD2}, and the \textit{CD2}-associated genes, including \textit{CD48}, \textit{CD53}, \textit{CD58}, and \textit{CD84}.  \textit{CD48}, a member of the CD2 subfamily of the immunoglobulin superfamily, is found on the surface of lymphocytes and other immune cells and participates in activation and differentiation pathways in these cells. \textit{CD53}, another member of the tetraspanin superfamily, regulates various cellular processes such as adhesion, migration, signaling, and cell fusion. \textit{CD58}, also known as lymphocyte function-associated antigen 3 (LFA-3), is a cell adhesion molecule that strengthens the adhesion and recognition between T cells and antigen-presenting cells, facilitating signal transduction necessary for an immune response. Lastly, \textit{CD84}, a member of the signaling lymphocyte activation molecule (SLAM) family, forms homophilic dimers by self-association and is reported as an important survival receptor in chronic lymphocytic leukemia. By studying the coactivation of these molecules, we hope to gain a deeper understanding of the complex interactions and signaling pathways involved in immune regulation. This could lead to the development of more effective therapeutic strategies for various immune-related diseases.}

We will investigate four pathways regarding the co-occurrence of mutational activation, including:
\begin{enumerate}
    \item \textbf{Pathway 1}: \textit{CTLA4}-activation stand-alone.
    \item \textbf{Pathway 2}: \textit{CTLA4-CD8A-CD8B} activated simultaneously.
    \item \textbf{Pathway 3}: \textit{CTLA4-CD2} activated simultaneously.
    \item \textbf{Pathway 4}: \textit{CTLA4-CD2-CD48-CD53-CD58-CD84} activated simultaneously.
\end{enumerate}
Advanced ML or Quantum AI has yet to study these mutational activation pathways to extend our knowledge. We summarize the biological meaning of quantified targets in \textbf{Appendix}~\ref{app:biological_exp}.

\subsubsection{Datasets}
We use The Cancer Genome Atlas (TCGA\cite{weinstein2013cancer}) with RNA expression data and Copy Number of Variation (CNV). We score $2,048$ genetic biomarkers based on its expression, which is equivalent to $2^{11}$ dimensional embedding generated by $11$-qubit system (\textbf{Section}~\ref{sec:pseudo-code}). The evaluated cohort includes $9,136$ patients, considered Big Data in the context of a cancer genetic study. The expression set $\bm{X}$ is normalized using Min-Max normalization, and the mutational signals are created from CNV with $\bm{Y} = 1$ if CNV $\neq 0$, otherwise $\bm{Y} = 0$. The evaluated expressions are continuous values, while the activation signals are binary.

\subsubsection{Experimental Settings}
All experiments were carried out using Python $3.7.0$, numpy $1.21.5$, sci-kit-learn $1.0.2$, and PyTorch $1.11$ on an Intel i9 processor (2.3 GHz, eight cores), 16GB DDR4 memory and GeForce GTX $1060$ Mobile GPU with 6GB memory. Besides, we made our implementation available at: \textcolor{blue}{\url{https://github.com/namnguyen0510/Biomarker-Discovery-with-Quantum-Neural-Networks}}; and all experimental history available at \textcolor{blue}{\url{https://tinyurl.com/55x77w4h}}. We train our sampler for $600$ trials with paralleled computing using ten CPU workers for each pathway.

\subsection{Quantum AI-driven Genetic Biomarkers for \textit{CTLA4}-activation Pathways}
\begin{figure}[t]
    \centering
    \includegraphics[width = \textwidth]{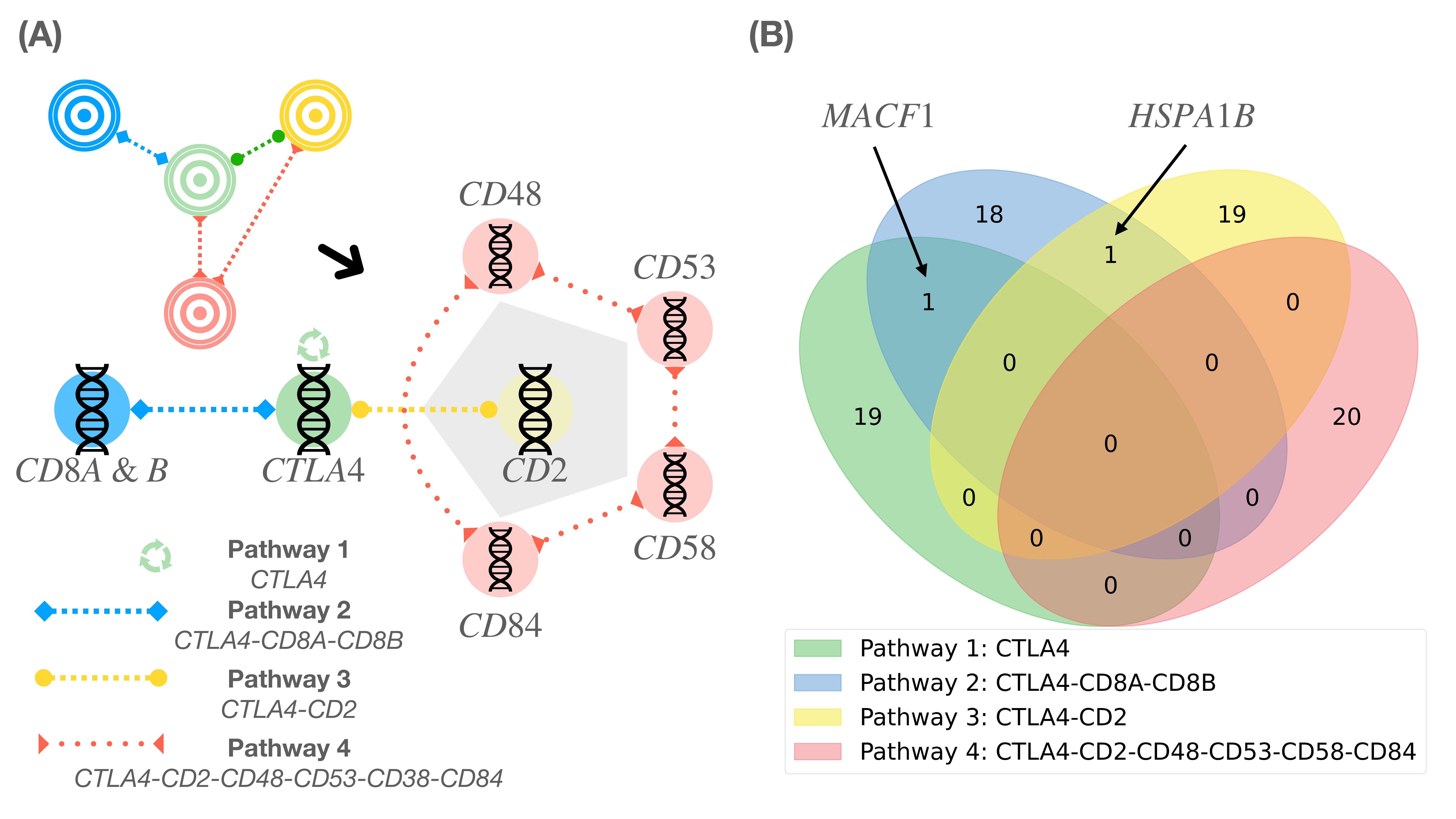}
    \caption{\textbf{(A)} \textbf{Activation Pathways Analyzed in Our Case-study:} We expanded the scope of immunological research by studying the coactivation of CTLA4, CD2, and associated genes, including CD48, CD53, CD58, and CD84. These molecules play significant roles in T-cell immune regulation, and their interactions could potentially lead to the development of more effective therapeutic strategies for various immune-related diseases. \textbf{(B)} \textbf{Venn Diagram of Discovered-Genetic Biomarker Sets regarding The Quantified Targeted Pathways:} \textit{MACF1}, a protein facilitating actin-microtubule interactions at the cell periphery, is a common genetic biomarker for both \textit{CTLA4} and \textit{CTLA4-CD8A-CD8B} pathways. On the other hand, \textit{HSPA1B}, a member of the heat shock protein 70 family that stabilizes existing proteins against aggregation, is the common genetic biomarker for \textit{CTLA4-CD8A-CD8B} and \textit{CTLA4-CD2} pathways. These findings highlight the potential of these biomarkers in understanding immune regulation and developing therapeutic strategies, which has not yet been well-studied, discussed in Section~\ref{sec:casestudy_ctla4}.}
    \label{fig:result_1}
\end{figure}
We summarize the case-study of \textit{CTLA4}-activation pathways in \textbf{Figure}~\ref{fig:result_1}(A). The full reports for each pathway is given in \textbf{SuppFig}~\ref{fig:result_pathway1}, ~\ref{fig:result_pathway2}, ~\ref{fig:result_pathway3} and \ref{fig:result_pathway4} in \textbf{Appendix}~\ref{app:sup_result}. We only consider the top 20 genetic biomarkers for further analysis. Of note, the discovered biomarker sets are distinctive for each studied pathway. Specifically, only one genetic biomarker \textit{MACF1} is founded as the genetic biomarker for \textit{CTLA4} and \textit{CTLA4-CD8A-CD8B} activation pathways (\textbf{Figure}~\ref{fig:result_1}(B)). Similarly, \textit{HSPA1B} is addressed as a common genetic biomarker for the pathways \textit{CTLA4-CD8A-CD8B} and \textit{CTLA4-CD2}. Apart from \textit{MACF1} and \textit{HSPA1B}, the remaining genetic biomarkers are distinctively associated with the quantified targets.

\subsection{Convergence Analysis}
We construct an end-to-end explainable quantum AI, illustrated in the results in  \textbf{SuppFig}~\ref{fig:result_pathway1}, ~\ref{fig:result_pathway2}, ~\ref{fig:result_pathway3} and \ref{fig:result_pathway4}. Of note, we are addressing a complex problem beyond the capacity of classical computers; thus, our proposed algorithm will likely suggest a sub-optimal solution. We show in these results of \textbf{SuppFig}~\ref{fig:result_pathway1}, ~\ref{fig:result_pathway2}, ~\ref{fig:result_pathway3} and \ref{fig:result_pathway4} that the proposed algorithm can effectively score genes and sampling biomarker sets as the sampling loss is reducing. Training beyond $600$ trials does not guarantee better solutions as most of the best loss is found by trial $400^{th}$. We average top-50$\%$ models to have a more robust inference of GSCORE$^\copyright$, which shows that higher GSCORE$^\copyright$ tends to have higher score variation since the markers are more frequently sampled by the quantum model. However, the score variation is extremely small with under $200\mu=2\times 10^{-6}$, indicating the well-convergence of neural solutions.

Furthermore, the output panels in \textbf{SuppFig}~\ref{fig:result_pathway1}, ~\ref{fig:result_pathway2}, ~\ref{fig:result_pathway3} and ~\ref{fig:result_pathway4} also report the landscape of hyper-optimization protocol, in which model configuration with the lower score is in a darker color (purple) and model configuration with a higher score is in brighter color (yellow). This analysis significantly reduces the cost of model deployment on actual quantum computers, as we can select the optimal design of quantum ansatz circuits based on analytical solutions using classical simulations.

\section{Significance of Discovered Genetic Biomarkers}\label{sec:cli_sig}
\subsection{Statistical Significance}\label{sec:stat_sig}
Using the cBioPortal\cite{cerami2012cbio} databases from nine projects\cite{zehir2017mutational, robinson2017integrative,miao2018genomic,hyman2018her,samstein2019tumor,rosen2020trk,bolton2020cancer,wu2022landscape,nguyen2022genomic} with a total of $73,717$ samples, we validated the statistical significance of the top-5 genetic biomarkers for each pathway (see Section~\ref{sec:data_aval}). We found that only a small proportion of the total samples contained mutations in certain biomarkers, with the highest mutated biomarkers being \textit{PAK1} ($1.5\%, n=858$) and \textit{RNF213} ($0.3\%, n=62$). However, the remaining biomarkers accounted for an extremely small proportion of all mutations, making it insufficient to use mutational profiles alone to study the relationship between these biomarkers and targets. We also found co-occurrence of mutations in certain biomarkers with significant statistical evidence regarding the \textit{CTLA4}-activation pathway. Specifically, \textit{UBA1}, \textit{HYOU1}, and \textit{RNF213} mutations were co-occurring in this pathway. Additionally, mutations in \textit{MAFC1}, \textit{WLS}, \textit{PSMD8}, and \textit{PAK1} were co-occurring in the \textit{CTLA4-CD8A-CD8B} pathways, while mutations in \textit{NAPA}, \textit{PGAP3}, \textit{CLIC4}, and \textit{PPM1G} were co-occurring in the activation of \textit{CTLA4-CD}2. Lastly, the extended activation pathway four was associated only with the coactivation of \textit{LCN2} and \textit{FAM107A} mutations.

\subsection{Clinical Significance}\label{sec:cli_sig}
We perform literature mining from the PubMed.gov library regarding $19$ top-5 genetic biomarkers, excluding CPE due to similar abbreviations (see Section~\ref{sec:data_aval}). The analysis is on publications over the three years from $2020$ to $2023$, up to 12/05/2023. \textbf{SuppTab}~\ref{tab:lit_mine} shows that $12$ over $19$ biomarkers are rarely known in clinical-associated literature but addressed as significant genetic biomarkers by our proposed model.

\subsubsection{Pathway 1: \textit{CTLA4}}
Hypoxia upregulated protein 1 is a protein that in humans is encoded by the \textit{HYOU1} gene. The protein encoded by this gene belongs to the heat shock protein 70 family. This gene uses alternative transcription start sites. Using a type of bone-forming cell called MC3T3-E1, the study\cite{zhou2020melatonin} demonstrated that high levels of glucose decrease the ability of the cells to survive and cause them to undergo programmed cell death. The high glucose levels also cause endoplasmic reticulum stress (ERS) by increasing the movement of calcium and producing a protein called binding immunoglobulin protein (BiP) in the endoplasmic reticulum. This results in the activation of eukaryotic initiation factor 2 alpha (eIF2alpha) downstream of a protein called PKR-like ER kinase (PERK). This, in turn, leads to the activation of a transcription factor called \textit{ATF4} and an increase in the production of a protein called C/EBP-homologous protein (CHOP), which is involved in the regulation of apoptosis in response to ER stress, as well as other proteins like \textit{DNAJC3}, \textit{HYOU1}, and \textit{CALR}. Besides, the discovered \textit{HYOU1} and \textit{HSPA1A} (in the \textit{HSPA1B} family) and \textit{DNAJB11, CALR, ERP29, GANAB, HSP90B1, HSPA5, LMAN1, PDIA4} and \textit{TXNDC5} were involved in the endoplasmic reticulum (ER) stress\cite{de2020proteins}. The analysis of how proteins interact with each other revealed certain genes, such as \textit{PTBP1}, \textit{NUP98}, and \textit{HYOU1}, that are linked to breast cancer brain metastasis\cite{an2021comprehensive}. \textit{HYOU1} plays a role in supporting the growth, spreading, and metabolic activity of papillary thyroid cancer by increasing the stability of \textit{LDHB} mRNA\cite{wang2021hyou1}. 

Apart from its significant protective function in the formation and progression of tumors, \textit{HYOU1} can be a promising target for treating cancer. It may be an immune-stimulating additive because it can trigger an antitumor immune response. Additionally, it can be a molecular target for treating various endoplasmic reticulum-related ailments\cite{rao2021biological}. The study\cite{lee2021expression} found that the secretion of certain substances in response to a communication between lung cancer cells and endothelial cells (ECs) led to an increase in the expression of \textit{HYOU1} in lung cancer spheroids. Additionally, direct interaction between ECs and lung cancer cells caused an upregulation of \textit{HYOU1} in multicellular tumor spheroids (MCTSs). When inhibiting \textit{HYOU1} expression, it reduced the malignant behavior and stemness of the cancer cells, facilitated apoptosis, and made the MCTSs more sensitive to chemotherapy drugs in lung cancer.

\textit{ETS2} is responsible for producing a transcription factor that controls the activity of genes related to both development and apoptosis. The protein it produces is not only a proto-oncogene but has also been found to play a role in regulating telomerase. A non-functional copy of this gene, known as a pseudogene, is also on the X chromosome. Due to alternative splicing, various transcript variants of this gene are generated. The transcription factor \textit{ETS2} controls the expression of genes responsible for various biological processes such as development, differentiation, angiogenesis, proliferation, and apoptosis. The transcription factor \textit{ETS2} has been shown to downregulate the expression of cytokine genes in resting T-cells. \textcolor{blue}{The research \cite{davoulou2020transcription} have investigated whether \textit{ETS2} also regulates the expression of lymphotropic factors (LFs) that are involved in T-cell activation/differentiation and the kinase \textit{CDK10}, which controls Ets-2 degradation and repression activity.} In vitro experiments demonstrated that Ets-2 overexpression increased the expression of certain LFs while decreasing \textit{CDK10} levels in both stimulated and unstimulated T-cells. \textcolor{blue}{Cyclin-dependent kinase 10 (CDK10) is a serine/threonine kinase related to CDC2 and plays a crucial role in various cellular processes such as cell proliferation, regulation of transcription, and cell cycle regulation. CDK10 has been identified as a candidate tumor suppressor in hepatocellular carcinoma, biliary tract cancers, and gastric cancer, but as a candidate oncogene in colorectal cancer (CRC)\cite{bazzi2021cdk10}. A study on CDK10's role in colorectal cancer revealed that it enhances cell growth, reduces chemosensitivity, and inhibits apoptosis by increasing the expression of \textit{BCL-2}\cite{weiswald2017inactivation}. This effect depends on its kinase activity, as colorectal cancer cell lines with a kinase-defective mutation exhibit an exaggerated apoptotic response and reduced proliferating capacity.CDK10 is a serine/threonine kinase that regulates various cellular processes. It is a candidate tumor suppressor in hepatocellular carcinoma, biliary tract cancers, and gastric cancer but an oncogene in colorectal cancer (CRC). CDK10 promotes cell growth, reduces chemosensitivity, and inhibits apoptosis by increasing the expression of \textit{BCL-2} in colorectal cancer. Kinase-defective mutations exaggerate apoptotic response and reduce proliferating capacity. The relation between the identified-genetic biomarker \textit{ETS2} and \textit{CDK10} could be used to develop new therapeutic applications of cancer treatment.} 


The relationship between \textit{CELF1} and \textit{ETS2} in colorectal cancer (CRC) and chemoresistance to oxaliplatin (L-OHP) is studied in \cite{wang2020rna}. \textit{CELF1} was overexpressed in human CRC tissues and positively correlated with \textit{ETS2} expression. Overexpression of \textit{CELF1} increased CRC cell proliferation, migration, invasion, and L-OHP resistance, while knockdown of \textit{CELF1} improved the response of CRC cells to L-OHP. Similarly, overexpression of \textit{ETS2} increased malignant behavior and L-OHP resistance in CRC cells. The study concluded that \textit{CELF1} regulates \textit{ETS2}, resulting in CRC tumorigenesis and L-OHP resistance, and may be a promising target for overcoming chemoresistance in CRC.

\textit{GPR116} or \textit{ADGRF5} is the probable G-Protein coupled receptor 116. \textit{GPR116} has been reported to be involved in cancer progression and predicts poor prognosis in other types of cancer. The study\cite{zheng2021gpr116} shows that \textit{GPR116} expression is upregulated in gastric cancer (GC) tissues and is positively correlated with tumor invasion and poor prognosis. \textit{GPR116} may be a novel prognostic marker and a potential therapeutic target for GC treatment. mRNA and protein expression of \textit{GPR116} in GC tissues and found that it was significantly upregulated, positively correlated with tumor node metastasis (TNM) staging and tumor invasion, and contributed to poor overall survival in GC patients\cite{kang2022expression}. \textit{GPR116} overexpression was also found to be an independent prognostic indicator in GC patients. Enrichment analysis revealed that \textit{GPR116} co-expression genes were mainly involved in various pathways. 

The effect of \textit{GPR116} receptor on NK cells concerning pancreatic cancer is studied in\cite{guo2023gpr116}, which found that \textit{GPR116} mice were able to efficiently eliminate pancreatic cancer through enhancing the proportion and function of NK cells in the tumor. The expression of \textit{GPR116} receptor decreased upon NK cells' activation, and \textit{GPR116} with NK cells showed higher cytotoxicity and antitumor activity in vitro and in vivo. Downregulation of \textit{GPR116} receptor also promoted the antitumor activity of NKG2D-CAR-NK92 cells against pancreatic cancer. These findings suggest that downregulation of \textit{GPR116} receptor could enhance the antitumor efficiency of CAR NK cell therapy.

\subsubsection{Pathway 2: \textit{CTLA4-CD8A-CD8B}}
The 26S proteasome (\textit{PSMD2}) is a complex enzyme comprising a 20S core and a 19S regulator arranged in a precise structure. The 20S core consists of four rings with 28 different subunits. Two rings contain seven alpha subunits each, while the others contain seven beta subunits each. The 19S regulator consists of a base with six ATPase subunits, two non-ATPase subunits, and a lid with up to ten non-ATPase subunits. Proteasomes are found in high concentrations throughout eukaryotic cells and break down peptides through an ATP/ubiquitin-dependent process outside lysosomes. The immunoproteasome, a modified version, plays a critical role in processing class I MHC peptides. This gene codes for one of the non-ATPase subunits in the lid of the 19S regulator. Besides its involvement in proteasome function, this subunit may also participate in the TNF signaling pathway as it interacts with the tumor necrosis factor type 1 receptor. A non-functional copy of this gene has been found on chromosome 1. Multiple transcript variants of this gene are produced through alternative splicing. \textit{PSMD2} and \textit{PSMD8} were significantly over-expressed in bladder urothelial carcinoma (BLCA) more than other cancers\cite{salah2021prognostic}. Besides, \textit{PSMD8} with \textit{AUNIP}, \textit{FANCI}, \textit{LASP1}, and \textit{XPO5} are potential targets for the creation of an mRNA vaccine to combat mesothelioma\cite{zhang2022identification}.

\textit{PAK1} is responsible for producing a member of the PAK protein family, which are serine/threonine p21-activating kinases. PAK proteins connect RhoGTPases to reorganize the cytoskeleton and nuclear signaling. They act as targets for small GTP-binding proteins such as CDC42 and RAC. This particular family member specifically regulates the movement and shape of cells. Different isoforms of this gene have been identified through alternative splicing, resulting in various transcript variants.

Regarding its mechanism, ipomoea batatas polysaccharides specifically encourage the degradation of \textit{PAK1} through ubiquitination and inhibit its downstream Akt1/mTOR signaling pathway, thereby resulting in an increased level of autophagic flux\cite{bu2021therapeutic}. \textit{PAK1} is a serine/threonine kinase gene overexpressed in some human breast carcinomas with poor prognosis, and aberrant \textit{PAK1} expression is an early event in the development of some breast cancers\cite{duderstadt2021chemical}. The structure of the actin cytoskeleton and protrusions in SW620 cells is related to their ability to move. Ce6-PDT treatment inhibits the migration of SW620 cells by reducing the activity of the \textit{RAC1}/\textit{PAK1}/\textit{LIMK1}/cofilin signaling pathway, and this inhibition is improved by decreasing the expression of the \textit{RAC1} gene\cite{wufuer2021downregulation}.

The increased presence of \textit{WLS} (Wnt Ligand Secretion Mediato) is an important indication of a negative outcome in breast cancer. It may have a vital function in the hormone receptor-positive (HR+) subtype\cite{zheng2020wntless}. Reducing the expression of \textit{SNHG17} in lung adenocarcinoma cells hindered cell growth, migration, and invasion while increasing apoptosis. \textit{SNHG17} acted as a sponge for miR-485-5p, resulting in increased expression of WLS. Therefore, \textit{SNHG17} accelerates lung 
adenocarcinoma progression by upregulating \textit{WLS} expression through sponging miR-485-5p\cite{li2020snhg17}. 

\textit{NDUFS5} belongs to the NADH dehydrogenase (ubiquinone) iron-sulfur protein family. The protein it encodes is a component of the NADH - ubiquinone oxidoreductase (complex I), the initial enzyme complex in the electron transport chain situated in the inner membrane of mitochondria. Through alternative splicing, multiple transcript variants of this gene are generated. Additionally, pseudogenes of this gene have been discovered on chromosomes 1, 4, and 17.

Machine learning algorithms have been used to classify some cancer types, but not lung adenocarcinoma, in which \textit{NDUFS5, P2RY2, PRPF18, CCL24, ZNF813, MYL6, FLJ41941, POU5F1B, and SUV420H1} were associated with alive without disease\cite{deng2020classify}. The study identified \textit{MACF1} with \textit{FTSJ3}, \textit{STAT1}, \textit{STX2}, \textit{CDX2} and \textit{RASSF4} that can be used as a signature to predict the overall survival of pancreatic cancer patients\cite{yang2020construction}. The poor response of low-grade serous ovarian carcinoma (LGSOC) to chemotherapy calls for a thorough genomic analysis to identify new treatment options. This study, 71 LGSOC samples were analyzed for 127 candidate genes using whole exome sequencing and immunohistochemistry to assess key protein expression. Mutations in \textit{KRAS}, \textit{BRAF}, and \textit{NRAS} genes were found in 47\% of cases. Several new genetic biomarkers were identified, including \textit{USP9X}, \textit{MACF1}, \textit{ARID1A}, \textit{NF2}, \textit{DOT1L}, and \textit{ASH1L}\cite{cheasley2021genomic}. To improve the treatment of glioblastomas and enhance patient survival, \textit{MACF1} can be used as a specific diagnostic marker that enhances the effectiveness of radiation therapy while minimizing damage to normal tissues\cite{bonner2020inhibition}. This approach could potentially lead to the development of new combination radiation therapies that target translational regulatory processes, which are often involved in poor patient outcomes. Another study presented data indicating that reduced \textit{MACF1} expression inhibited melanoma metastasis in mice by blocking the epithelial-to-mesenchymal transition process. Therefore, \textit{MACF1} could be a potential target for melanoma treatment\cite{wang2020decreasing}. 

\textit{MACF1} is responsible for producing a substantial protein that consists of multiple spectrin and leucine-rich repeat (LRR) domains. The encoded protein belongs to a family of proteins that connect various cytoskeletal components. Specifically, this protein plays a role in enabling interactions between actin and microtubules at the outer edges of cells, and it also connects the microtubule network to cellular junctions. Multiple transcript variants of this gene are produced through alternative splicing, although the complete structure of some of these variants has yet to be determined\cite{griffith2017civic}. A study included 695 patients with hepatocellular carcinoma (HCC), divided into a training group of 495 patients and a validation group of 200 patients\cite{huang2021development}. A nomogram was developed using T stage, age, and the mutation status of \textit{DOCK2}, \textit{EYS}, \textit{MACF1}, and \textit{TP53}. The nomogram was found to have good accuracy in predicting outcomes and was consistent with the actual data. The study also found that T-cell exclusion may be a potential mechanism for malignant progression in the high-risk group. In contrast, the low-risk group may benefit from immunotherapy and \textit{CTLA4} blocker treatment. In conclusion, the study developed a nomogram based on mutant genes and clinical parameters and identified the underlying association between these risk factors and immune-related processes.

\subsubsection{Pathway 3: \textit{CTLA4-CD2}}
Fucoxanthin is a natural pigment present in brown seaweeds, and its derivative, fucoxanthin (FxOH), has been shown to effectively induce apoptosis (programmed cell death) in various cancer cells. The role of Chloride intracellular channel 4 (\textit{CLIC4}), which plays a crucial role in cancer development and apoptosis, in FxOH-induced apoptosis was also investigated\cite{yokoyama2021effects}. Treatment with FxOH induced apoptosis in human CRC DLD-1 cells. FxOH treatment downregulated \textit{CLIC4}, integrin beta1, \textit{NHERF2}, and \textit{pSMAD2} (Ser(465/467)) compared to control cells, without affecting \textit{RAB35} expression. \textit{CLIC4} knockdown suppressed cell growth and apoptosis, and apoptosis induction by FxOH was reduced with \textit{CLIC4} knockdown. The expression levels of \textit{CLIC4} and \textit{GAS2L1} were found to be higher in circulating tumor cells (CTCs) from pancreatic cancer patients compared to peripheral blood mononuclear cells\cite{zhu2020gas2l1}. Besides, the overexpression of \textit{CLIC4} was associated with unfavorable outcomes in multiple cohorts of CN-AML patients\cite{huang2020aberrant}. 

The role of actin-binding proteins, including profiling, fascin, and ezrin, in the metastasis of non-small cell lung cancer (NSCLC)\cite{kolegova2020increases}. The study collected tumor and adjacent normal lung tissue samples from 46 NSCLC patients and used real-time PCR and Western blotting to determine the levels of \textit{PFN1}, \textit{FSCN1}, and \textit{EZR} mRNAs and proteins. The results showed that patients with lymphatic metastasis had higher expression levels of the profilin, fascin, and ezrin mRNAs and profilin and fascin proteins. In contrast, mRNA and protein expression levels increased in patients with distant metastasis. The activation of AKT signaling in the progression of colorectal cancer (CRC) can be influenced by the lncHCP5/miR-299-3p/\textit{PFN1}\cite{bai2020knockdown}. The loss of \textit{PFN1} leads to the activation of several signaling pathways, including AKT, NF-(k)B, and WNT. On the other hand, overexpression of PFN1 in cells with high levels of SH3BGRL can counteract SH3BGRL-induced metastasis and tumor growth by upregulating \textit{PTEN} and inhibiting the PI3K-AKT pathway\cite{zhang2021adaptor}.

\textit{PPM1G} codes for a large protein that contains spectrin and leucine-rich repeat (LRR) domains. It belongs to a family of proteins that act as bridges between different cytoskeletal elements. Specifically, this protein facilitates the interaction between actin and microtubules at the periphery of cells and links the microtubule network to cellular junctions. The level of \textit{PPM1G} expression in LIHC may be influenced by promoter methylation, CNVs, and kinases and could be linked to immune infiltration. High \textit{PPM1G} expression was found to be related to mRNA splicing and the cell cycle according to GO terms. These findings suggest that \textit{PPM1G} could be a prognostic indicator for liver hepatocellular carcinoma patients and may play a role in the tumor immune microenvironment\cite{lin2021prognostic}. Besides, the irc-\textit{PGAP3} plays a significant role in the growth and advancement of triple-negative breast cancer (TNBC), thus making it a potential target for the treatment of TNBC patients.

\subsubsection{Pathway 4: \textit{CTLA4-CD2-CD48-CD53-CD58-CD84}}
\textit{MT1G} (Metallothionein 1G) and \textit{MT1H} have the potential to suppress tumor growth and are regulated by DNA methylation in their promoter regions. In addition, they are associated with serum copper levels and may be linked to the survival rate of patients with hepatocellular carcinoma\cite{udali2021trace}. Besides, \textit{MT1G}, \textit{CXCL8}, \textit{IL1B}, \textit{CXCL5}, \textit{CXCL11}, and \textit{GZMB} are over-expressed in colorectal cancer tissues compared to normal tissues\cite{meng2020identification}. Three genes (\textit{SLC7A11}, \textit{HMOX1}, and \textit{MT1G}) were identified as differentially expressed genes (DEGs) associated with renal cancer prognosis using survival analysis screening\cite{chen2022functions}. \textit{SLC7A11} and \textit{HMOX1} were found to be upregulated in renal cancer tissues, while \textit{MT1G} was downregulated. The combination of receiver operating characteristic (ROC) curves, Kaplan-Meier analysis, and Cox regression analysis revealed that high expression of \textit{SLC7A11} was a prognostic risk factor for four different types of renal cancers, low expression of \textit{HMOX1} was a poor prognostic marker for patients, and increased expression of \textit{MT1G} increased the prognostic risk for three additional classes of renal cancer patients, except for those with renal papillary cell carcinoma.

\textit{FAM107A} (Family With Sequence Similarity 107 Member A) with \textit{ADAM12}, \textit{CEP55}, \textit{LRFN4}, \textit{INHBA}, \textit{ADH1B}, \textit{DPT}, and \textit{LOC100506388} were analyzed and evaluated as potential prognostic genes for gastric cancer\cite{huang2021comprehensive}. Among these genes, \textit{LRFN4}, \textit{DPT}, and \textit{LOC100506388} were identified as having a potential prognostic role in gastric cancer, as determined through a nomogram. Besides, the interaction pairs of \textit{HCG22/EGOT-hsa-miR-1275-FAM107A} and \textit{HCG22/EGOT-hsa-miR-1246}-Glycerol-3-phosphate dehydrogenase 1 are likely to have a significant role in laryngeal squamous cell carcinoma.

\textit{LCN2} produces a protein classified as a lipocalin family member. Lipocalins are known for their ability to transport small hydrophobic molecules like lipids, steroid hormones, and retinoids. The specific protein encoded by this gene is called neutrophil gelatinase-associated lipocalin (NGAL), and it plays a significant role in innate immunity. NGAL sequesters iron-containing siderophores, which helps limit bacterial growth and infection\cite{griffith2017civic}. Besides, \textit{LCN2} is an innate immune protein that regulates immune responses by promoting sterile inflammation. \textit{LCN2} is a biomarker associated with radioresistance and recurrence in nasopharyngeal carcinoma (NPC)\cite{zhang2021lcn2}. \textit{LCN2} expression was upregulated in radioresistant NPC tissues and associated with NPC recurrence. Knocking down \textit{LCN2} enhances the radiosensitivity of NPC cells, while ectopic expression of \textit{LCN2} confers additional radioresistance. \textit{LCN2} may interact with \textit{HIF-1A} and facilitate the development of a radioresistant phenotype. \textit{LCN2} is a promising target for predicting and overcoming radioresistance in NPC. Moreover, the downregulation of the immune response, influenced by specific metastasis-evaluation genes (\textit{BAMBI}, \textit{F13A1}, \textit{LCN2}) and their associated immune-prognostic genes (\textit{SLIT2}, \textit{CDKN2A}, \textit{CLU}), was found to increase the risk of post-operative recurrence\cite{luo2021development}. Higher \textit{LCN2} expression was associated with poor clinical outcomes and correlated with increased infiltration of various immune cells. \textit{LCN2} may serve as a genetic biomarker for immune infiltration and poor prognosis in cancers, suggesting potential therapeutic targets for cancer treatment\cite{xu2020integrative}.

\color{blue}
\section{Discussion}
\subsection{Complexity in Biomarker Identification for \textit{CTLA4} Activation Pathway}
Identifying the T cell gene CTLA4 and its genetic biomarker presents unique challenges, particularly when detecting certain molecular biomarkers expressed in the bloodstream. The detection of CTLA4 in T cells is complex due to several factors. First, CTLA4 is predominantly found in intracellular compartments before activation and only becomes increasingly detectable on the cell surface upon activation. This necessitates precise timing for effective detection. Second, the proportion of CTLA4-positive T-cell subgroups in the peripheral blood and tumor tissues could be higher, making their detection difficult and costly. Despite the challenges, identifying genetic biomarkers associated with CTLA4 has several advantages over detecting molecular biomarkers in the bloodstream. Genetic biomarkers can provide information about genetic susceptibility, genetic responses to environmental exposures, subclinical or clinical disease markers, or indicators of response to therapy\cite{chen2011biomarkers}. They can help identify high-risk individuals reliably and promptly so that they can either be treated before the onset of the disease or as soon as possible. When a genetic biomarker is identified in a cancer through molecular or genetic testing, it tells the physician what makes the cancer grow and thrive. That information allows physicians to decide the most effective treatment for the patient\cite{cancerGov}.

\subsection{Model Adaptation for Further Applications}
The further improvement of the proposed quantum neural network can consider the intricate nature of epigenetic modifications, which is of utmost importance. Epigenetic modifications are chemical alterations that occur to the DNA molecule itself or to the proteins tightly bound to it, and they play a crucial role in regulating gene expression and cellular differentiation. These modifications can include phosphorylation, acetylation, or methylation of various amino acids, and they not only add a layer of complexity but also enrich the genomic landscape with many potential biomarkers. To enhance the accuracy and reliability of our quantum neural network model, we need to integrate and analyze these diverse epigenetic markers. Doing so can unravel complex biological interactions and pathways associated with immune responses. This will allow us to identify novel and clinically relevant biomarkers for immunotherapy that may not be discovered through traditional methods.

Furthermore, addressing the complexities of epigenetic modifications will facilitate a more nuanced understanding of the model's adaptability. This understanding will offer insights into its potential applications and limitations in real-world clinical settings. By taking a comprehensive approach, we ensure that our model is robust and versatile, accommodating the vast array of biomarkers introduced by epigenetic modifications. In turn, this contributes to the advancement of personalized medicine in immuno-therapy, where we can tailor treatments to an individual's unique genetic makeup, epigenetic modifications, and immune system response.
\color{black}

\section{Conclusion}\label{sec:conclusion}
To this end, a new framework based on the quantum AI model, used for genetic biomarker discovery in biomedical research (\textbf{Section}~\ref{sec:method}), has been introduced. The proof of concept is demonstrated in four targeted pathways associating with therapeutic-target \textit{CTLA4} in \textbf{Section}~\ref{sec:results}. Our model found clinical-relevant and notably potential biomarkers/targets for cancer treatment, which are extensively validated through statistical methods and literature mining (\textbf{Section}~\ref{sec:cli_sig}).

We suggest several research directions that can be extended from the study. First, deploying the proposed quantum AI models on real quantum computers is worth investigating in the future, in which the effect of noise should be addressed toward the model's efficiency and effectiveness. Second, extension to other pathway activation is possible as the proposed algorithm is generalized. Finally, in \textit{vivo} and in \textit{vitro}, validations of the discovered in \textit{silico} biomarkers will translate the findings to therapeutic solutions for cancer treatment and prevention.

\section{Declarations}

\subsection{Abbreviations}
\begin{table}[h]
    \centering
    \begin{tabular}{|c|c|}
        \toprule
        \textbf{Abbreviation} & \textbf{Meaning}  \\
        \toprule
        AI & Artificial Intelligence\\
        ML & Machine Learning\\
        QNN & Quantum Neural Networks\\
        QML & Quantum Machine Learning\\
        mRMR & Maximum Relevance - Minimum Redundancy\\
        GSCORE$\copyright$ & Gene Score\\
        temp & Temperature\\
        SMBO & Sequential Model-based Optimization\\
        CNV & Copy Number Variation\\
        TCGA & The Cancer Genome Atlats\\
        Genetic Biomarker(s) & Biomarker(s) identified from genomic data\\
        \bottomrule
    \end{tabular}
    \caption{Abbreviations Table}
\end{table}

\subsection{Ethics approval and consent to participate}
Not applicable

\subsection{Consent for publication}
Not applicable

\subsection{Availability of data and materials}\label{sec:data_aval}
\begin{itemize}
    \item The implementation is made open source at: \url{https://github.com/namnguyen0510/Biomarker-Discovery-with-Quantum-Neural-Networks}.
    \item The processed data and the experimental history are available at: \url{https://drive.google.com/drive/folders/1QsL0N5E2xppWK2pmvfLI621PeXV2OQUA?usp=share_link}.
    \item The evaluation for statistical significance (Section~\ref{sec:stat_sig}) by cBioPortal is reported at: \url{https://drive.google.com/drive/folders/1ubbLEg35Pz9i_Wjr4uKvMuS1KNw5rrsS?usp=sharing}.
    \item The literature mining for clinical significance (Section~\ref{sec:cli_sig}) is reported at: \url{https://drive.google.com/drive/folders/1NHDuJQoah8ZmlP8zWIPD6aYwwWTxqoYe?usp=sharing}.
\end{itemize}

\subsection{Competing Interests}
The author declares that they have no known competing financial interests reported in this paper.

\subsection{Funding}
Not applicable

\subsection{Author's Contribution:} Phuong-Nam Nguyen conceptualized the algorithm and performed numerical analysis. 

\subsection{Acknowledgements}
The author would like to thank colleagues for stimulating discussion.

\subsection{Authors' Information}
Phuong-Nam Nguyen is a current Ph.D. student in Electrical Engineering at the University of South Florida. He earned a Bachelor's degree in Mathematics and Education from the Hanoi University of Education and a Master's in Statistics from the University of South Florida. He is currently a lecturer and research fellow at Phenikaa University. His research interests are applying classical and quantum machine intelligence to biological problems, including oncology research, molecular dynamics modeling, large genome data analysis, and biomarker identification.

\bibliography{output}
\bibliographystyle{abbrv}

\newpage
\begin{appendices}
\setcounter{figure}{0}
\renewcommand{\figurename}{SuppFig.}
\setcounter{table}{0}
\renewcommand{\tablename}{SuppTab.}

\section{mRMR Criteria in The Context of Genetic Biomarker Discovery}\label{app:mrmr}
An efficient and well-known method for genetic biomarker identification as feature selection problems is forward stage-wise search with \textit{relevancy} and \textit{redundancy} criteria\cite{peng2005feature}. The technique aims to simultaneously maximize the relevancy between a new feature $\bm{X}_i$ with target $\bm{Y}$
\begin{equation}\label{equa:rel}
    \text{REL}(\bm{X}_i,\bm{Y}) := I(\bm{X}_i,\bm{Y})
\end{equation}
while 
minimizing the redundancy of the chosen set $\mathbb{S}$\cite{nguyen2014effective}
\begin{equation}\label{equa:red}
    \text{RED}(\bm{X}_i|\mathbb{S}) := \sum_{\bm{X}_j \in \mathcal{S}} I(\bm{X}_i,\bm{X}_j).
\end{equation}
Note that $I(\bm{X}_i,\bm{X}_j)$ is pair-wise mutual information computed from the feature set. The optimization problem is given as
\begin{equation}
    \max_{\bm{X}_i \in \mathcal{X} \setminus \mathbb{S}} = \{\text{REL}(\bm{X}_i)-\text{RED}(\bm{X}_i|\mathbb{S}) \}.
\end{equation}
This problem is equivalent to minimizing the loss value:
\begin{equation}
    \mathcal{L} = I(\bm{X}_i, \bm{Y}) - \lambda \sum_{\bm{X}_j \in \mathbb{S}} I(\bm{X}_i, \bm{X}_j),
\end{equation}
where $\lambda = 1/|\mathbb{S}|$ is considered weights that scales the pairwise-mutual information between $\bm{X}_i$ and $\bm{X}_j$. 

\section{Quantum Unitary Transformation as Representations}\label{app:qnn}
We will present the quantum unitary transformation in QNNs as a representation.
\begin{definition}
    A representation of a group $\mathcal{G}$ is a homomorphism $\Phi: \mathcal{G} \rightarrow GL(V)$  with $GL(V):= \{f \in End(V)| f\textbf{ is invetible}\}$. We have $End(V)$ as the endomorphism ring of (finite-dimensional) vector space $V$, or a set of linear mapping from $V$ to $V$.
\end{definition}
Let
\begin{equation}
    \phi: \mathbb{Z}/n\mathbb{Z} \rightarrow \mathbb{C}^{*}
\end{equation}
where $\mathbb{Z}/n\mathbb{Z}$ is the integer modulo $n$ and
$$\phi([m]) = e^{2\pi i m/n}.$$ The mapping induced by the direct sum $\oplus$:
\begin{equation}
    \psi[m] = \begin{bmatrix}
        e^{\dfrac{2\pi m i}{n}} & 0\\
         0 & e^{-\dfrac{2\pi m i}{n}}
    \end{bmatrix}
\end{equation}
is equivalent to the representation $\mathbb{Z}/n\mathbb{Z} \rightarrow GL_2{\mathbb{C}}$:
\begin{equation}
    \phi[m] = \begin{bmatrix}
        \cos(\dfrac{2\pi m}{n}) & -\sin(\dfrac{2\pi m}{n})\\
         \sin(\dfrac{2\pi m}{n}) & \cos(\dfrac{2\pi m}{n})
    \end{bmatrix}
\end{equation}
Thus, the quantum unitary transformations in \textbf{Equation}~\ref{equa:rz} and ~\ref{equa:ry} are presented as the representation class $\psi[m]$ and $\phi[m]$, respectively. We note two main differences, which are (1) the flipped sign in $\psi[m]$ and $\bm{R}_Z$ and (2) the representation $\psi[m]$ and $\phi[m]$ intake rational values of $\dfrac{m}{n}$ while parameterized-rotation in \textbf{Equation}~\ref{equa:rz} and ~\ref{equa:ry} intake real value. In the first difference, considering $\psi[-m]$ addresses the issue. Regarding the latter difference, taking a large value of $n$ will make the representation asymptotic in the quantum unitary transformations.

By generalizing the core of quantum AI as mathematical concepts in representation theory, we have introduced a mathematical approach to discovery in cancer.

\section{Biological Meaning of Quantified Target Activation}\label{app:biological_exp}
\begin{table}[h]
  \centering
  \caption{\textbf{Summary of Target Genes with Biological Meaning from \cite{griffith2017civic}.}}
  \label{tab:target-genes}
  \begin{tabular}{|p{1cm}|p{14cm}|}
    \midrule
    \textbf{Target Gene} & \textbf{Summary} \\
    \midrule
    \textit{CTLA4} & This gene is a member of the immunoglobulin superfamily and encodes a protein that transmits an inhibitory signal to T cells. The protein contains a V domain, a transmembrane domain, and a cytoplasmic tail. Alternate transcriptional splice variants encoding different isoforms have been characterized. The membrane-bound isoform functions as a homodimer interconnected by a disulfide bond, while the soluble isoform functions as a monomer. Mutations in this gene have been associated with insulin-dependent diabetes mellitus, Graves disease, Hashimoto thyroiditis, celiac disease, systemic lupus erythematosus, thyroid-associated orbitopathy, and other autoimmune diseases. \\
    \midrule
    \textit{CD8A, CD8B} & The CD8 antigen is a cell surface glycoprotein found on most cytotoxic T lymphocytes that mediate efficient cell-cell interactions within the immune system. The CD8 antigen acts as a coreceptor with the T-cell receptor on the T lymphocyte to recognize antigens displayed by an antigen-presenting cell in class I MHC molecules. The coreceptor functions as either a homodimer composed of two alpha chains or a heterodimer composed of one alpha and one beta chain. Both alpha and beta chains share significant homology to immunoglobulin variable light chains. This gene encodes the CD8 alpha chain. Multiple transcript variants encoding different isoforms have been found for this gene. \\
    \midrule
    \textit{CD2} & A surface antigen of the human T-lymphocyte lineage is expressed on all peripheral blood T cells (summarized by Sewell et al., 1986 [PubMed 3490670]). It is one of the earliest T-cell markers on more than 95\% of thymocytes; it is also found on some natural killer cells but not on B lymphocytes. Monoclonal antibodies directed against CD2 inhibit the formation of rosettes with sheep erythrocytes, indicating that CD2 is the erythrocyte receptor or is closely associated with it. \\
    \midrule
    \textit{CD48} & This gene encodes a CD2 subfamily of immunoglobulin-like receptors member, which includes SLAM (signaling lymphocyte activation molecules) proteins. The encoded protein is found on the surface of lymphocytes and other immune cells, dendritic cells, and endothelial cells and participates in activation and differentiation pathways in these cells. The encoded protein does not have a transmembrane domain but is held at the cell surface by a GPI anchor via a C-terminal domain, which may be cleaved to yield a soluble receptor form. Multiple transcript variants encoding different isoforms have been found for this gene. \\
    \midrule
    \textit{CD53} & The protein encoded by this gene is a member of the transmembrane four superfamily, also known as the tetraspanin family. Most of these members are cell-surface proteins that are characterized by the presence of four hydrophobic domains. The proteins mediate signal transduction events that play a role in the regulation of cell development, activation, growth, and motility. This encoded protein is a cell surface glycoprotein known to be complex with integrins. It contributes to the transduction of CD2-generated signals in T cells and natural killer cells and has been suggested to play a role in growth regulation. Familial deficiency of this gene has been linked to an immunodeficiency associated with recurrent infectious diseases caused by bacteria, fungi, and viruses. Alternative splicing results in multiple transcript variants. \\
    \midrule
    \textit{CD58} & This gene encodes a member of the immunoglobulin superfamily. The encoded protein is a ligand of the T lymphocyte CD2 protein and functions in the adhesion and activation of T lymphocytes. The protein is localized to the plasma membrane. Alternatively, spliced transcript variants have been described. \\
    \midrule
    \textit{CD84} & This gene encodes a membrane glycoprotein member of the signaling lymphocyte activation molecule (SLAM) family. This family forms a subset of the larger CD2 cell-surface receptor Ig superfamily. The encoded protein is a homophilic adhesion molecule expressed in numerous immune cell types and regulates receptor-mediated signaling in those cells. Alternate splicing results in multiple transcript variants. \\
    \bottomrule
  \end{tabular}
\end{table}
We summarize the biological meaning of the quantified targets in \textbf{Table}~\ref{app:biological_exp}.

\section{Supplemental Result of Quantum AI-driven Biomarkers}\label{app:sup_result}

\begin{figure}
    \centering
    \includegraphics[width = \textwidth]{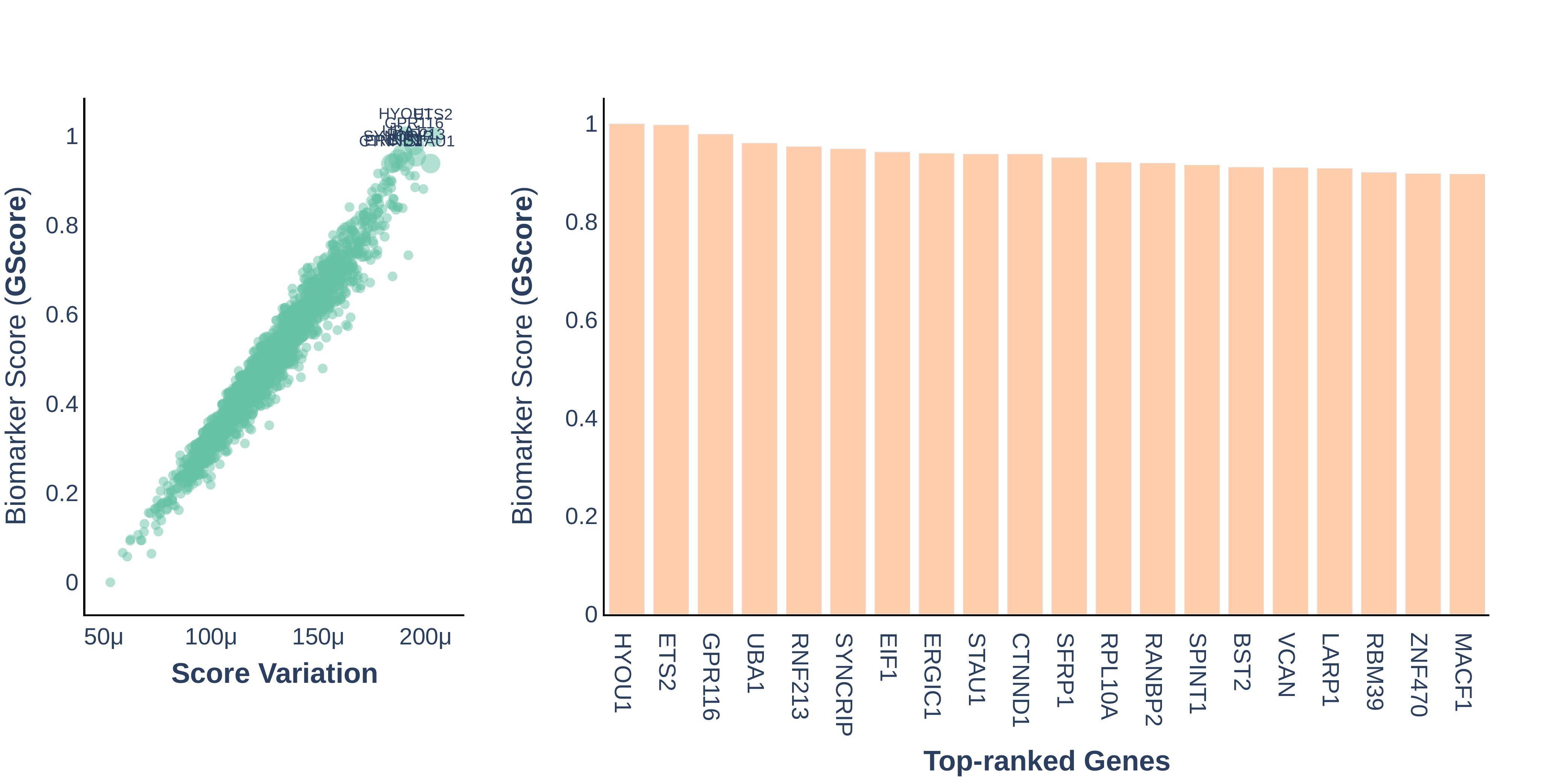}
    \includegraphics[width = 0.27\textwidth]{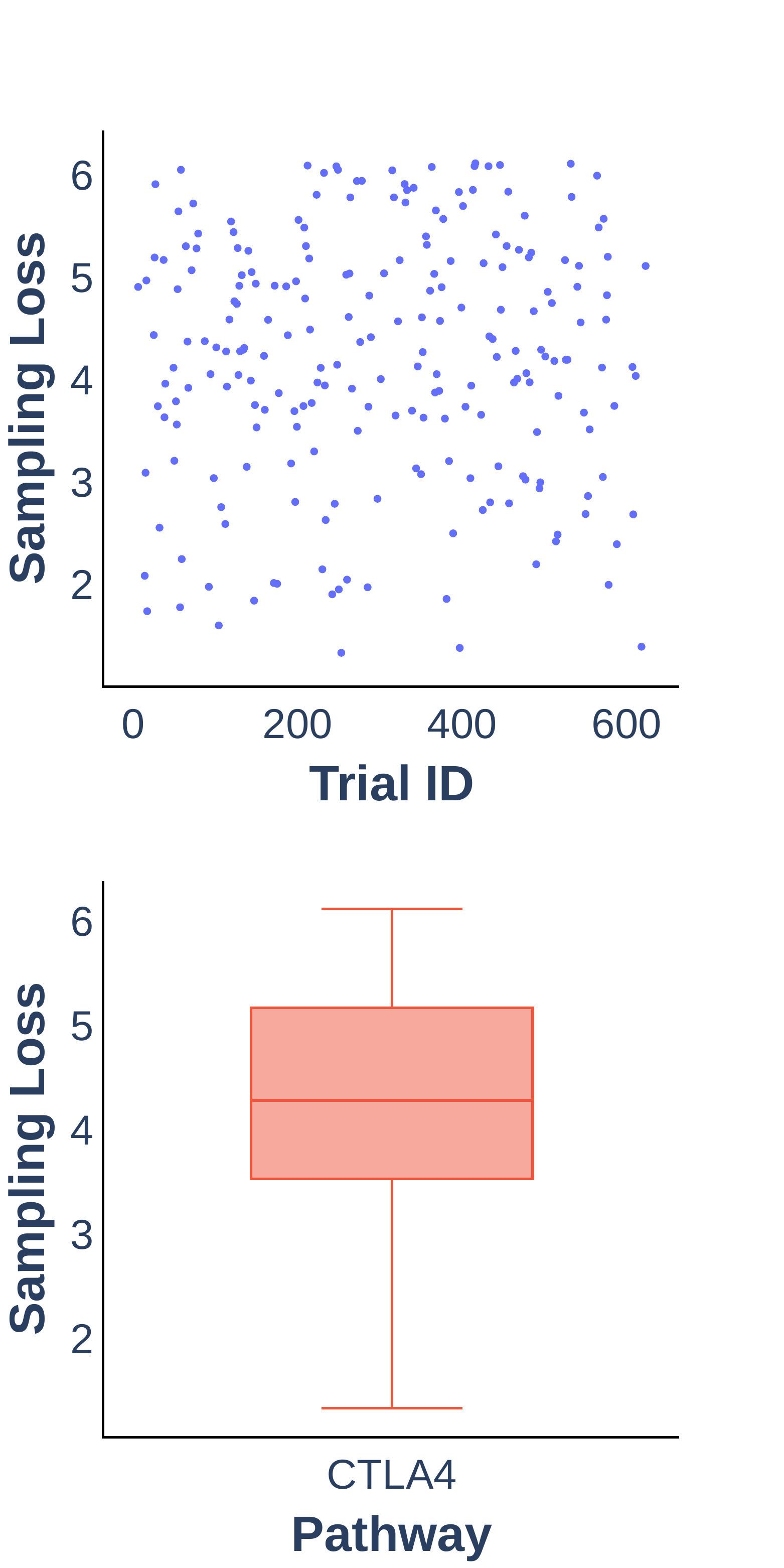}
    \includegraphics[width = 0.72\textwidth]{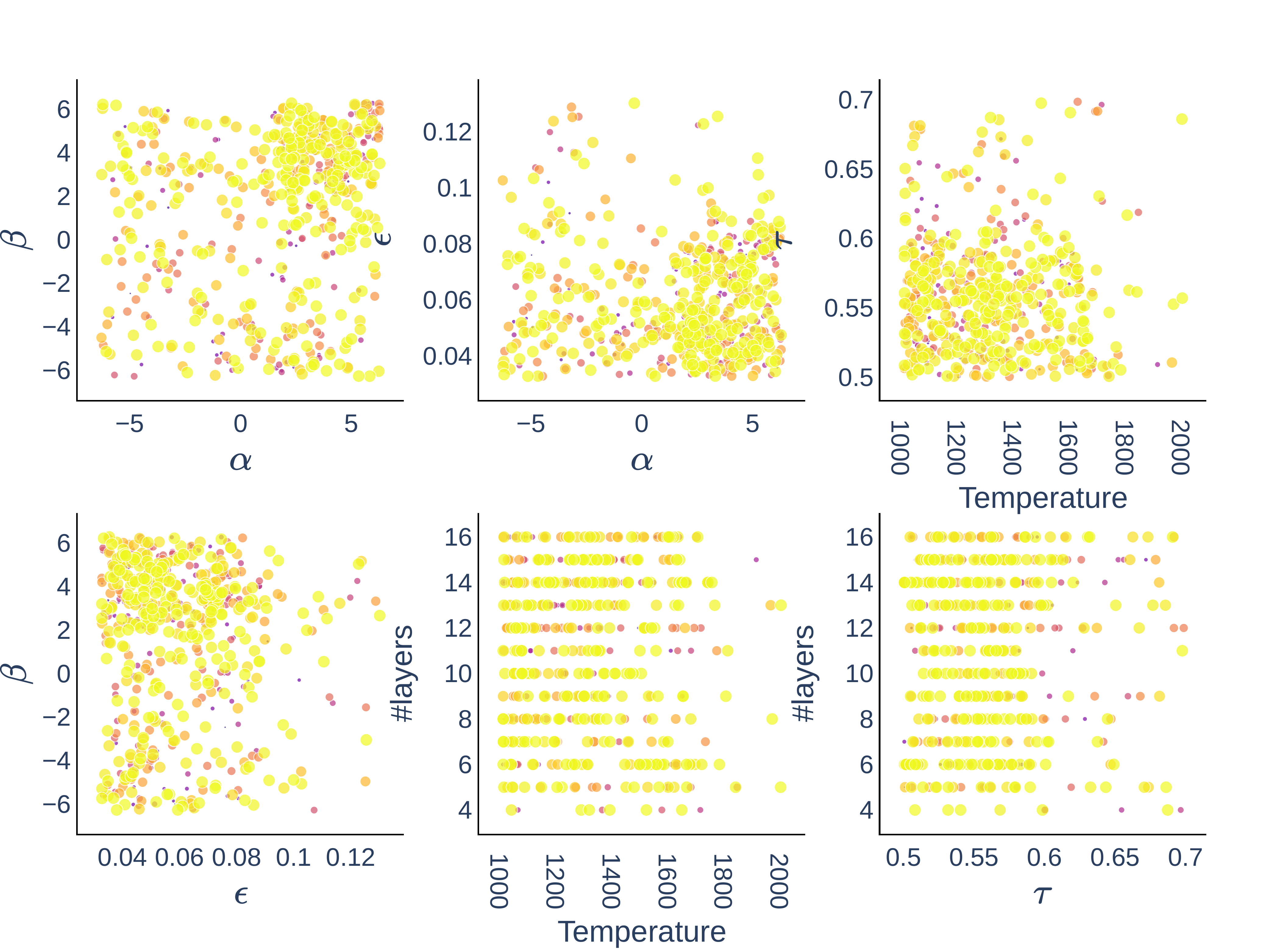}
    \caption{\textbf{Experimental Report of Quantum AI-driven Genetic Biomarkers Discovered for \textit{CTLA4} Pathway}. (Top) GSCORE$^\copyright$ using top-$50\%$ samplers. The neural solutions are well-converged as the score variation is under $200\mu, \mu = 10^{-6}$. (Bottom) Convergence analysis of quantum sampler. The model configuration with a lower score is in darker color (purple), and the model configuration with a higher score is in brighter color (yellow).}
    \label{fig:result_pathway1}
\end{figure}
\begin{figure}
    \centering
    \includegraphics[width = \textwidth]{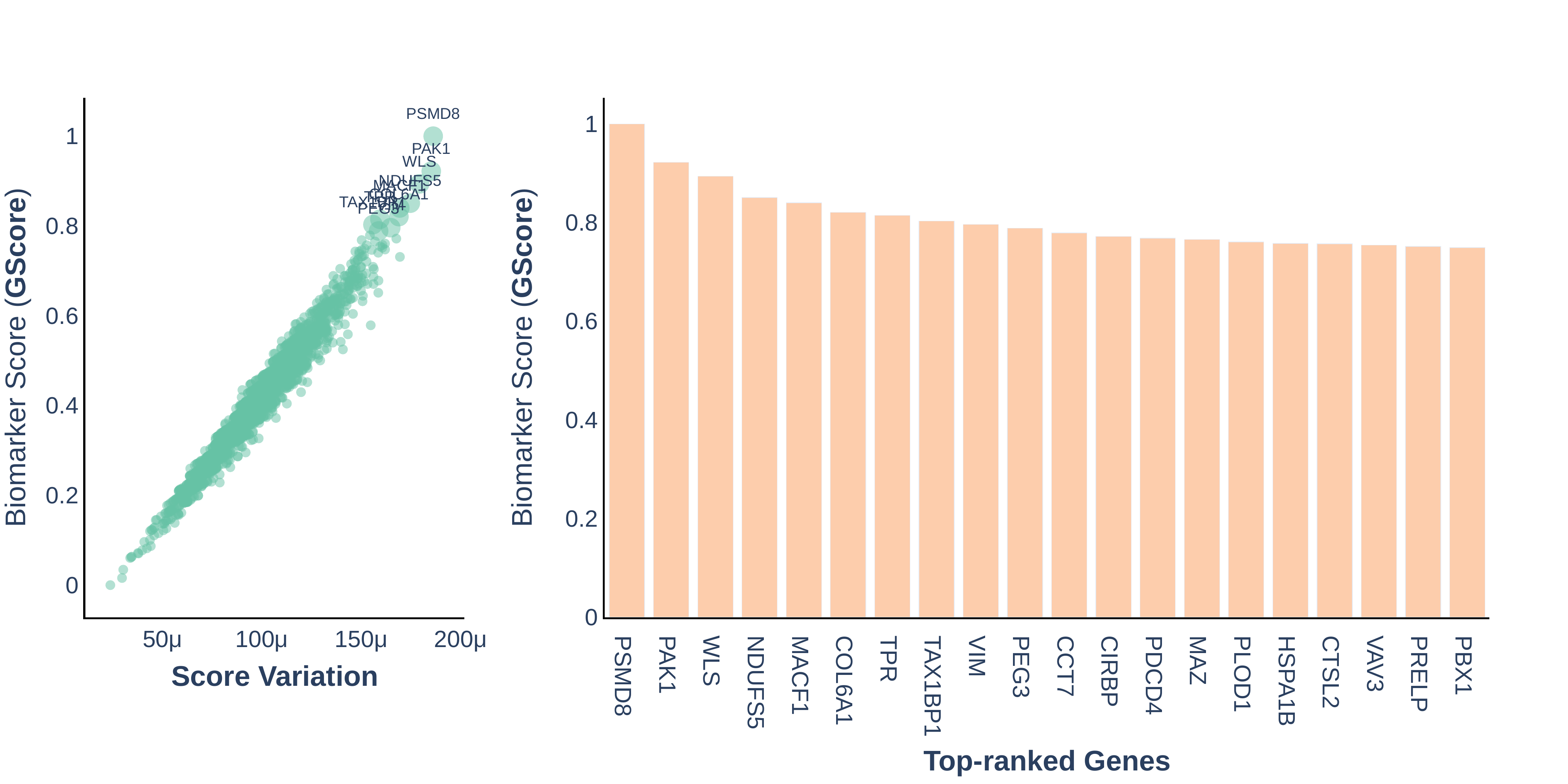}
    \includegraphics[width = 0.27\textwidth]{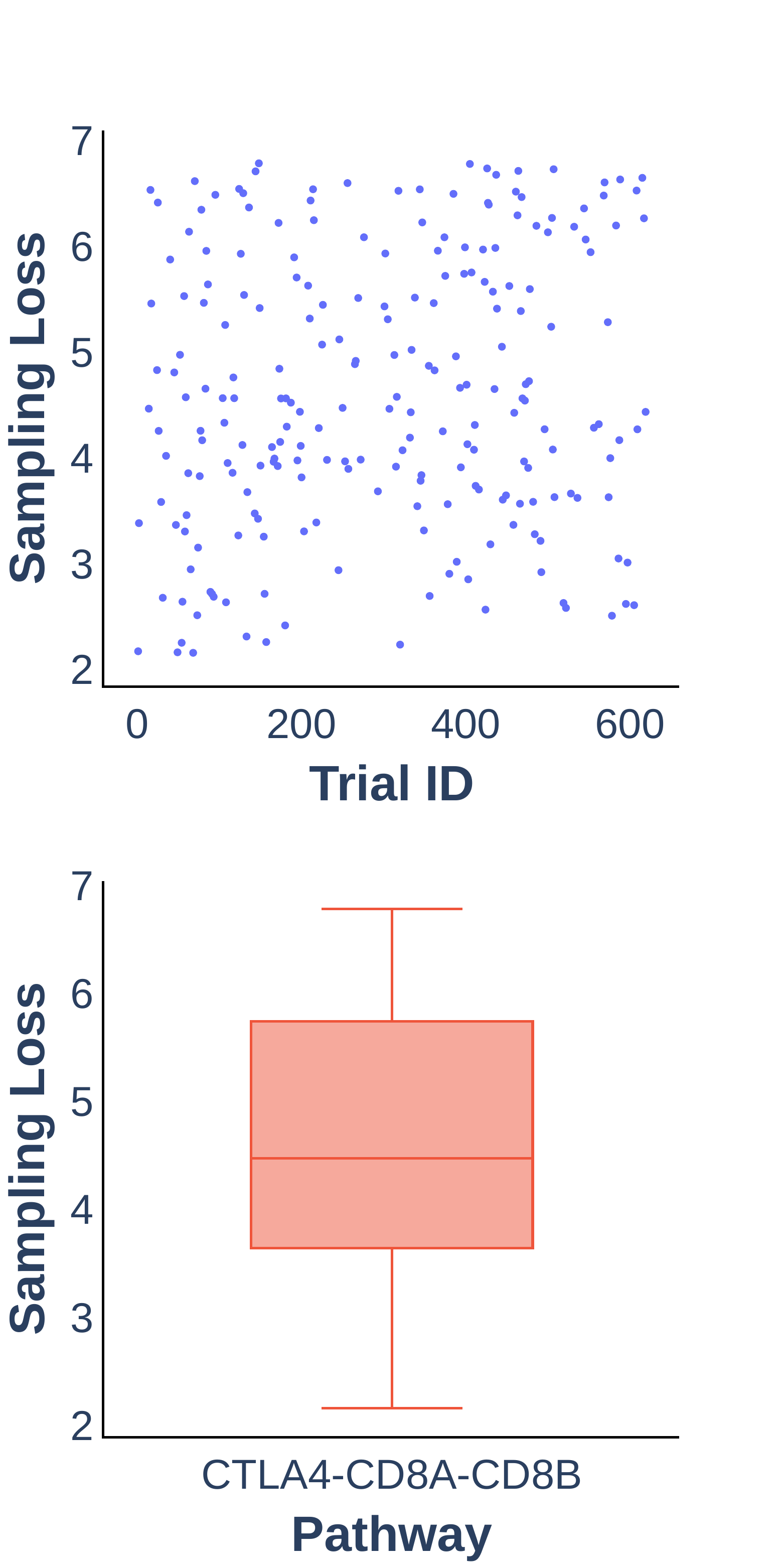}
    \includegraphics[width = 0.72\textwidth]{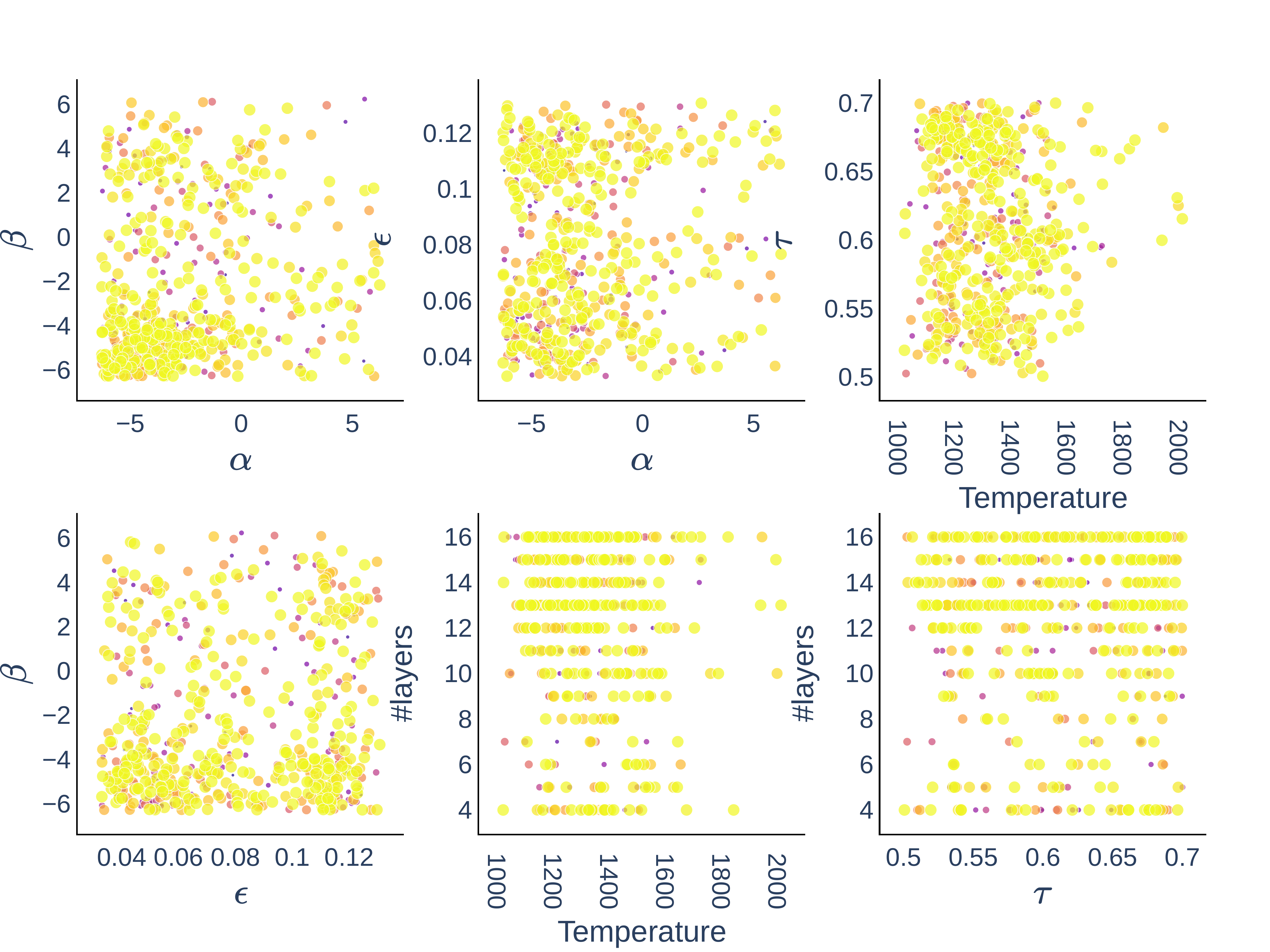}
    \caption{\textbf{Experimental Report of Quantum AI-driven Genetic Biomarkers Discovered for \textit{CTLA4-CD8A-CD8B} Pathway}. (Top) GSCORE$^\copyright$ using top-$50\%$ samplers. The neural solutions are well-converged as the score variation is under $200\mu, \mu = 10^{-6}$. (Bottom) Convergence analysis of quantum sampler. The model configuration with a lower score is in darker color (purple), and the model configuration with a higher score is in brighter color (yellow).}
    \label{fig:result_pathway2}
\end{figure}
\begin{figure}
    \centering
    \includegraphics[width = \textwidth]{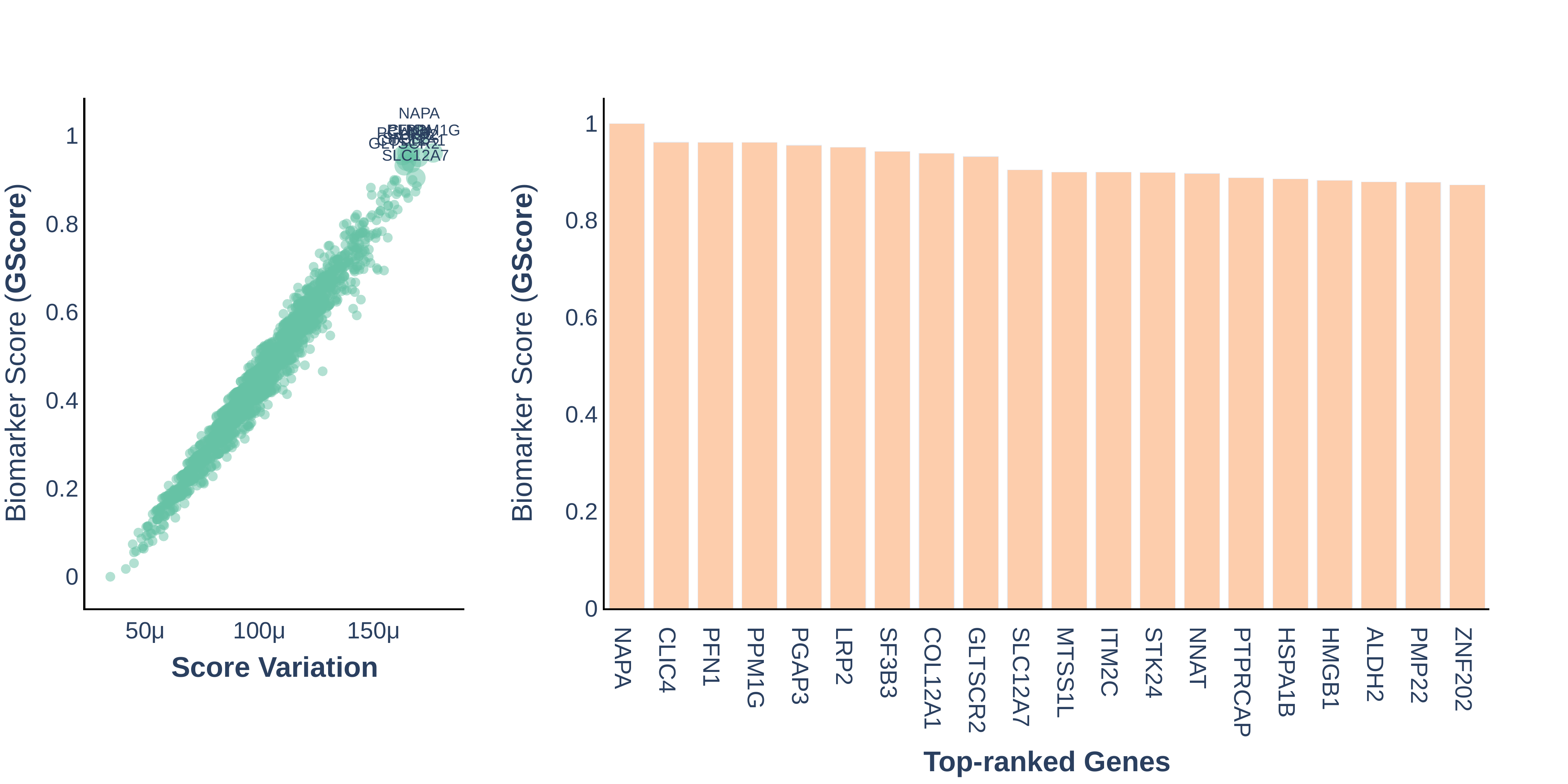}
    \includegraphics[width = 0.27\textwidth]{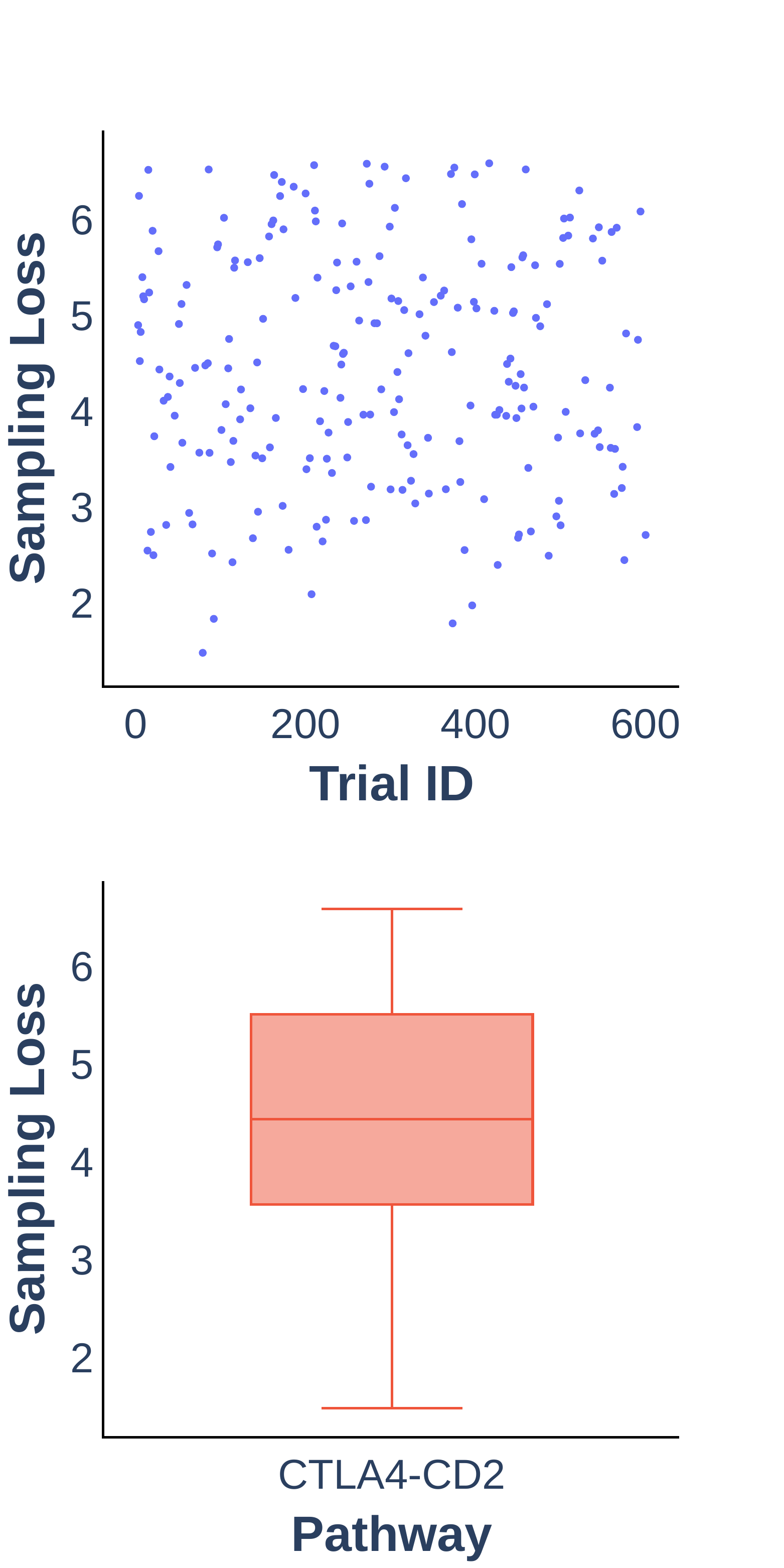}
    \includegraphics[width = 0.72\textwidth]{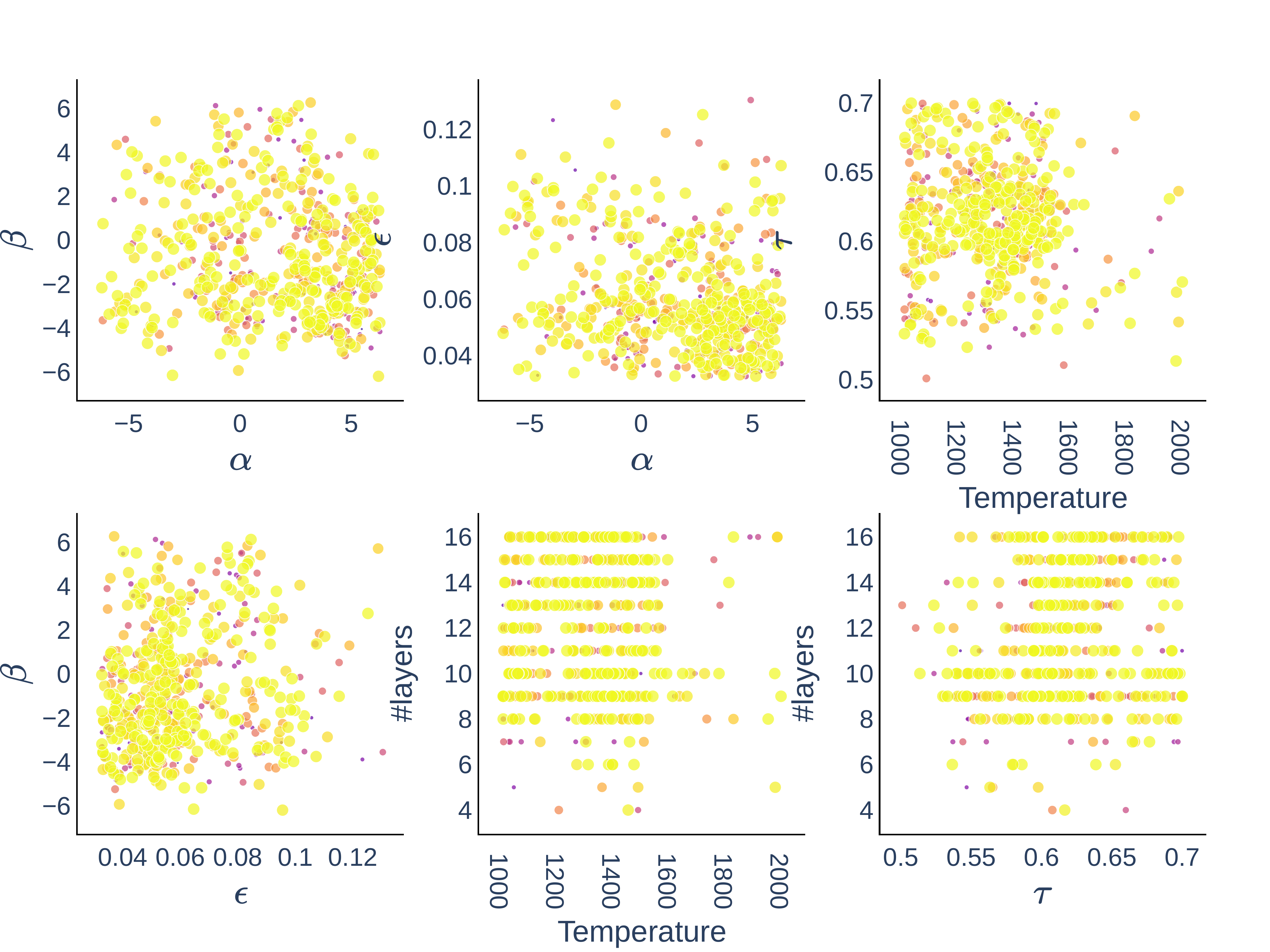}
    \caption{\textbf{Experimental Report of Quantum AI-driven Genetic Biomarkers Discovered for \textit{CTLA4-CD2} Pathway}. (Top) GSCORE$^\copyright$ using top-$50\%$ samplers. The neural solutions are well-converged as the score variation is under $200\mu, \mu = 10^{-6}$. (Bottom) Convergence analysis of quantum sampler. The model configuration with a lower score is in darker color (purple), and the model configuration with a higher score is in brighter color (yellow).}
    \label{fig:result_pathway3}
\end{figure}
\begin{figure}
    \centering
    \includegraphics[width = \textwidth]{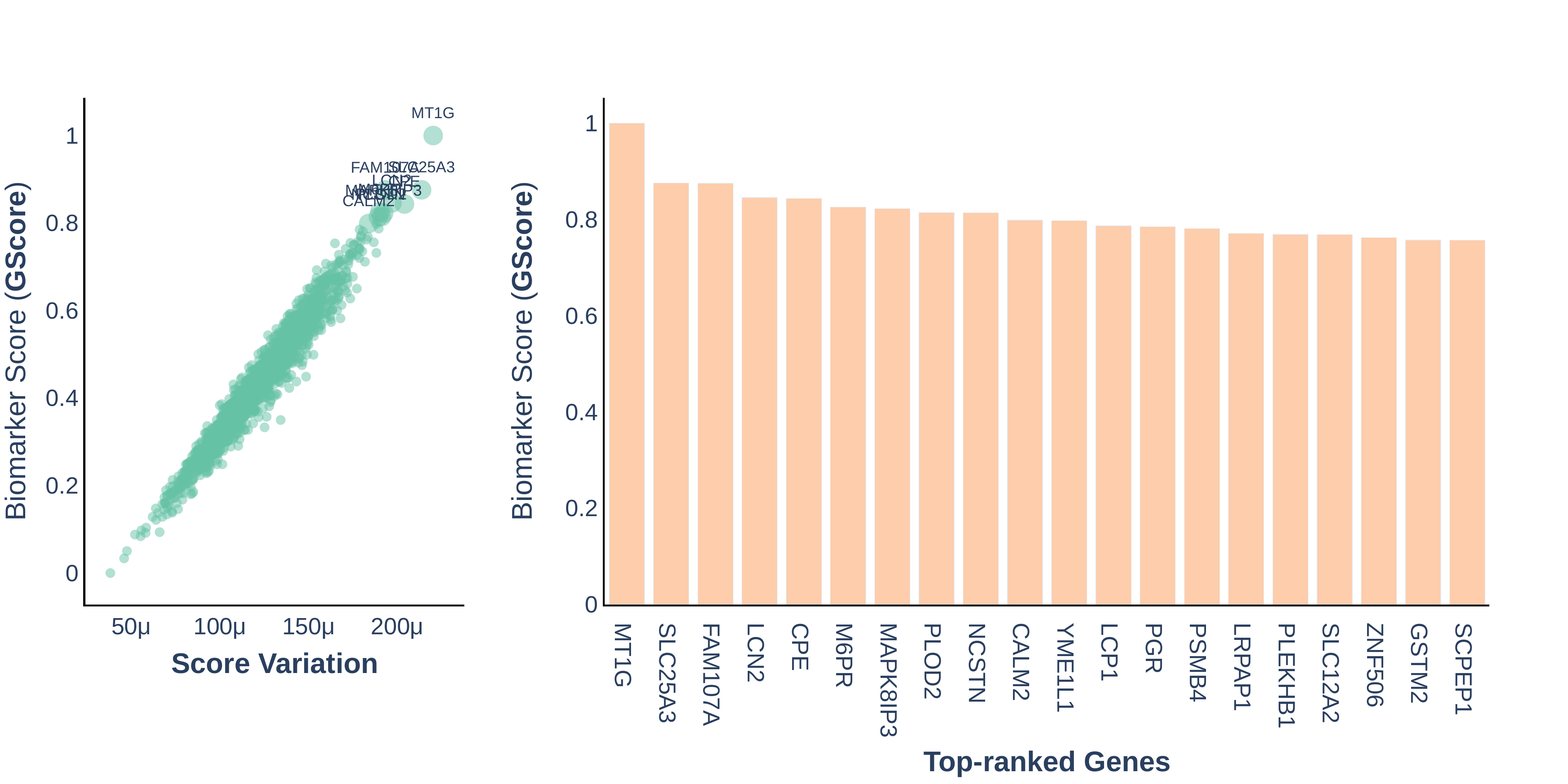}
    \includegraphics[width = 0.27\textwidth]{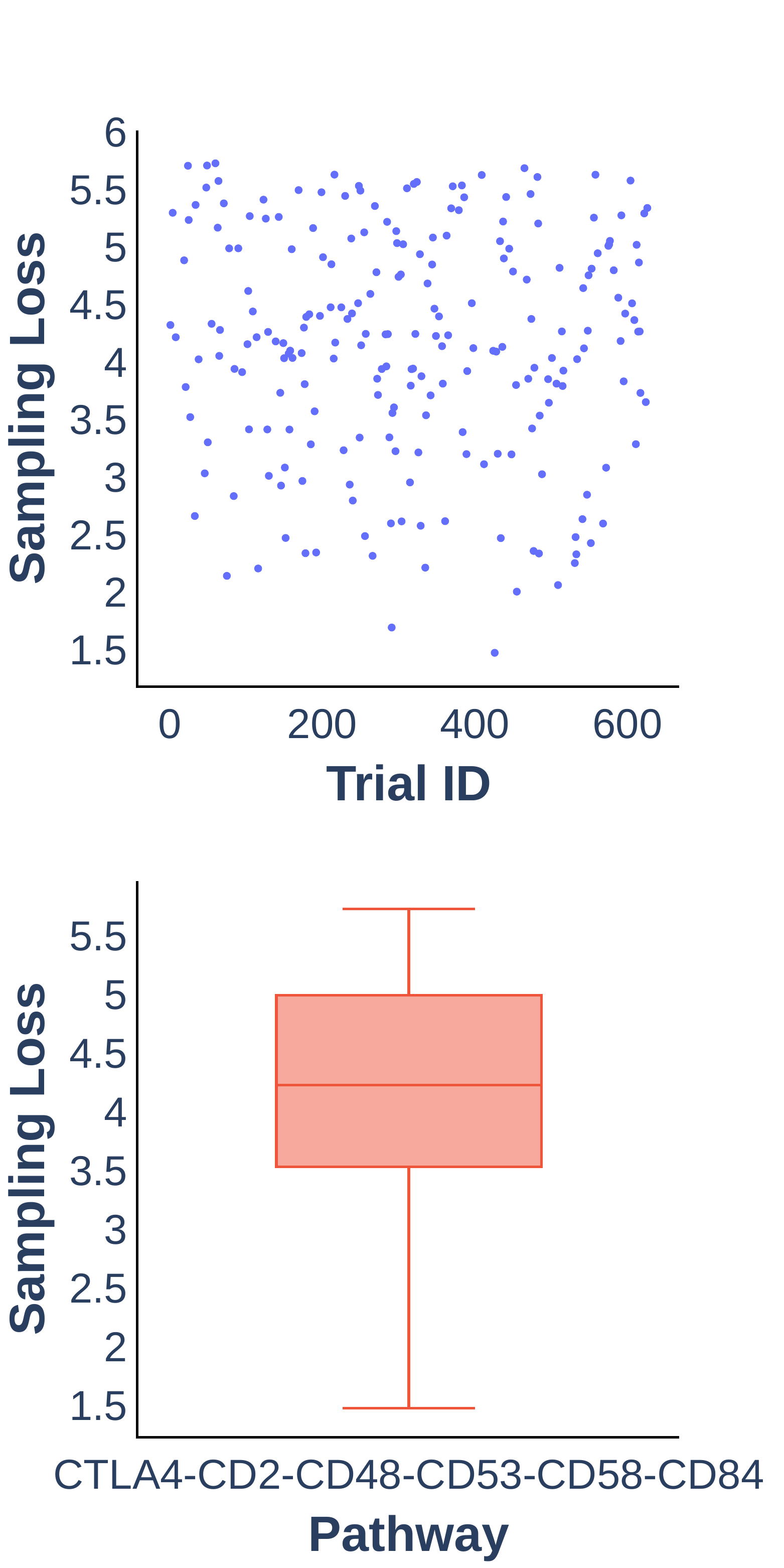}
    \includegraphics[width = 0.72\textwidth]{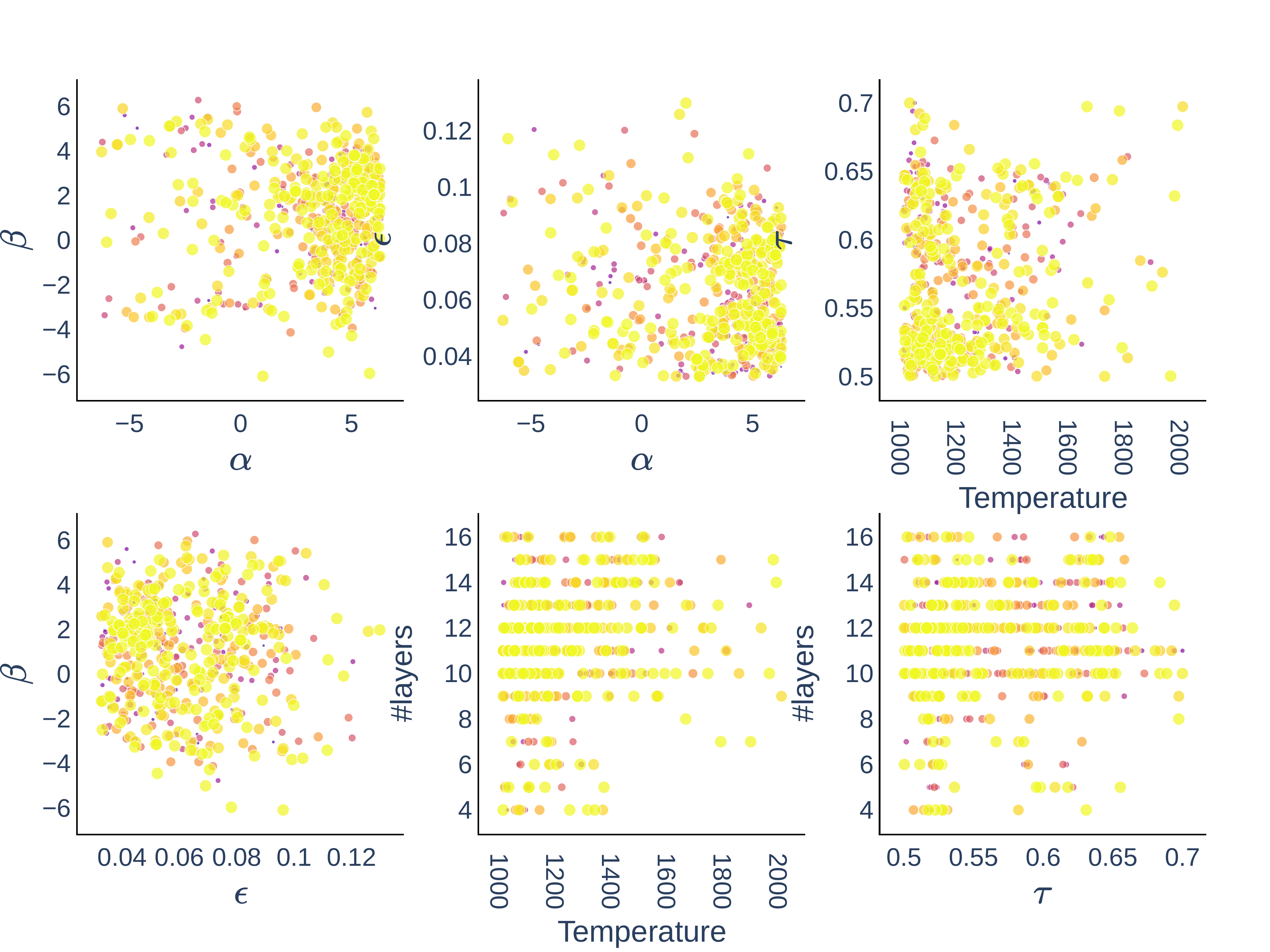}
    \caption{\textbf{Experimental Report of Quantum AI-driven Genetic Biomarkers Discovered for \textit{CTLA4-CD2-CD48-CD53-CD58-CD84} Pathway}. (Top) GSCORE$^\copyright$ using top-$50\%$ samplers. The neural solutions are well-converged as the score variation is under $200\mu, \mu = 10^{-6}$. (Bottom) Convergence analysis of quantum sampler. The model configuration with a lower score is in darker color (purple), and the model configuration with a higher score is in brighter color (yellow).}
    \label{fig:result_pathway4}
\end{figure}

\begin{table}[t]
    \centering
    \caption{\textbf{Literature Mining from \textit{PubMed.gov} Library of Top-5 Discovered Genetic Biomarkers.} Markers with * are rarely known in clinical literature, mentioned in under $100$ papers.}
    \begin{tabular}{|c|c|}
    \toprule
        \textbf{Query} & \textbf{Paper Count} \\
        \toprule
        \textit{SLC25A3}* & 14 \\
        \textit{NDUFS5}* & 8 \\
        \textit{PGAP3}* & 20 \\
        \textit{WLS} & 224 \\
        \textit{LCN2} & 743 \\
        \textit{RNF213} & 180 \\
        \textit{PFN1} & 110 \\
        \textit{PAK1} & 353 \\
        \textit{FAM107A}* & 21 \\
        \textit{PSMD8}* & 6 \\
        \textit{MACF1}* & 59 \\
        \textit{HYOU1}* & 37 \\
        \textit{ETS2}* & 92 \\
        \textit{PPM1G}* & 19 \\
        \textit{NAPA} & 397 \\
        \textit{UBA1} & 200 \\
        \textit{CLIC4}* & 51 \\
        \textit{GPR116}* & 15 \\
        \textit{MT1G}* & 41 \\
        \bottomrule
        \end{tabular}
    \label{tab:lit_mine}
\end{table}
We show the inference report of the proposed model in \textbf{SuppFig}~\ref{fig:result_pathway1}, ~\ref{fig:result_pathway2}, ~\ref{fig:result_pathway3} and ~\ref{fig:result_pathway4}.

\end{appendices}

\end{document}